\documentclass[final,times,twocolumn]{elsarticle}
\usepackage{graphicx,amssymb,amsmath}
\usepackage{multirow,dcolumn,bm,latexsym,soul}
\newcolumntype{K}[1]{>{\centering\arraybackslash}p{#1}}

\journal{Physics of the Dark Universe}
\begin{document}
\newcommand{\be}{\begin{equation}}
\newcommand{\ee}{\end{equation}}
\newcommand{\bq}{\begin{eqnarray}}
\newcommand{\eq}{\end{eqnarray}}

\begin{frontmatter}

\title{Consistency of local and astrophysical tests of the stability of fundamental constants}
\author[inst1,inst2]{C. J. A. P. Martins\corref{cor1}}\ead{Carlos.Martins@astro.up.pt}
\author[inst3,inst4]{M. Vila Mi\~nana}\ead{meritxellvila@josoc.cat}
\address[inst1]{Centro de Astrof\'{\i}sica da Universidade do Porto, Rua das Estrelas, 4150-762 Porto, Portugal}
\address[inst2]{Instituto de Astrof\'{\i}sica e Ci\^encias do Espa\c co, CAUP, Rua das Estrelas, 4150-762 Porto, Portugal}
\address[inst3]{Faculty of Physics, University of Barcelona, Carrer de Mart\'{\i} i Franqu\`es 1, 08028 Barcelona, Spain}
\address[inst4]{Faculty of Mathematics and Computer Science, University of Barcelona, Gran Via de les Corts Catalanes 585, 08007 Barcelona, Spain}
\cortext[cor1]{Corresponding author}

\begin{abstract}
Tests of the stability of nature's fundamental constants are one of the cornerstones of the ongoing search for the new physics which is required to explain the recent acceleration of the universe. The two main settings for these tests are high-resolution spectroscopy of astrophysical systems (mainly in low-density absorption clouds along the line of sight of bright quasars) and laboratory comparisons of pairs of atomic clocks. Here we use standard chi-square techniques to perform a global analysis of all currently available data, studying both the consistency of tests of stability of different constants (specifically the fine-structure constant $\alpha$, the proton-to-electron mass ratio $\mu$ and the proton gyromagnetic ratio $g_p$) and the consistency between local laboratory and astrophysical tests. We start by doing a model-independent analysis (studying the internal consistency of the various available datasets) but also explore specific phenomenological models motivated by string theory and grand unification. Overall there is weak (one to two sigma) evidence of variations, at the level of up to a few parts per million, and reasonable agreement between laboratory and astrophysical tests. This result holds even if one removes from the analysis the Webb {\it et al.} archival dataset of $\alpha$ measurements. Forthcoming astrophysical facilities, such as the ESPRESSO spectrograph, should be able to confirm or rule out these hints.
\end{abstract}


\end{frontmatter}

\section{Introduction} 

The observational evidence for the acceleration of the universe demonstrates that our canonical theories of cosmology and particle physics are incomplete: some yet unknown physical mechanism is needed to explain it \cite{SN1,SN2}. Current and future astrophysical facilities, together with local experiments, should therefore search for, identify and ultimately characterize this new physics. Tests of the stability of fundamental constants (or fundamental couplings) are one of the cornerstones of this search. The field has rapidly developed in the last few years, both due to some indications of possible detections of spacetime variations of the fine-structure constant $\alpha$ by Webb {\it et al.} \cite{Webb} and due to significant progress in the relevant astronomical instrumentation \cite{Moriond}. Two recent reviews can be found in \cite{Uzan,ROPP}.

Nature is characterized by a set of physical laws and of fundamental dimensionless couplings, and historically we have assumed that both of these are spacetime-invariant. While it is clear that for the former this is a cornerstone of the scientific method (it's hard to imagine how one could do science at all if it were not the case), one must also realize that for the latter this is only a convenient simplifying assumption. We have no 'theory of constants', and we don't know what role they play in physical theories. Our current definition of a fundamental constant is simply an operational one: any parameter whose value cannot be calculated within a given theory, but must be found experimentally.

Particle physics experiments have established beyond doubt that fundamental couplings \textit{run} with energy. Similarly, in many extensions of the standard model they will also \textit{roll} in time and \textit{ramble} in space (i.e., they will depend on the local environment). In particular, this will be the case in theories with additional spacetime dimensions, such as string theory. A detection of varying fundamental couplings will show that the Einstein Equivalence Principle is violated (and therefore that gravity can't be purely geometry), and that there is a fifth force of nature \cite{Carroll,Dvali,Chiba}. Tests of the stability of fundamental couplings are also relevant for dynamical dark energy models, as has been recently explored \cite{NunLid,Parkinson,Doran,Reconst,Pinho2}.

Direct high-resolution astrophysical spectroscopy measurements of $\alpha$ and the proton-to-electron mass ratio $\mu=m_p/m_e$ are, in most cases, carried out in the optical/ultraviolet (there are a few exceptions to this for the $\mu$ case), and up to redshifts now exceeding $z=4$. On the other hand, in the radio band, and typically at lower redshifts, one can measure various combinations of $\alpha$, $\mu$ and the proton gyromagnetic ratio $g_p$. (We note that throughout this work we use the term measurement in the generic sense, i.e. referring to both detections and null results.) On the other hand, laboratory tests with atomic clocks can measure the current drift ranges of these constants and/or combinations thereof. A compilation of astrophysical and local measurements available by early 2017 can be found in \cite{ROPP}.

In the present work we make use of this compilation, together with several other sensitive measurements that have been published more recently, and carry out a global analysis of all this data, using standard chi-square techniques. Our goals are both to check the consistency among different datasets and the consistency between local and astrophysical measurements, though for the latter one generally needs to assume a model. In passing we will also assess the constraining power of the datasets of dedicated measurements, by leaving the archival data of Webb {\it et al.} out of some of the analysis and comparing the constraints obtained from these dedicated measurements with those from the complete dataset of currently available measurements.

In principle, one could assume that any variations of $\alpha$, $\mu$ and $g_p$ might be independent of each other (as we do in the next section), and even be due to different physical mechanisms. However, if such variations do exist, the more parsimonious hypothesis is that they will be due do a single underlying mechanism. It follows that if a model for this mechanism is provided, the variations will not be independent of each other, though the relations between the variations will clearly be model-dependent. Therefore in the present work, in addition to a model-independent global analysis of the astrophysical data and (separately) of the atomic clock data, we will also relate the local and astrophysical measurements by assuming that the dynamical mechanism responsible for the variations is a scalar field \cite{Campbell,Damour}. Finally, in order to relate the variations of the three couplings in specific models we will rely on the phenomenological formalism developed in \cite{Coc,Luo}, for which we will consider three specific examples.

Our work builds upon and improves earlier analysis for atomic clock measurements \cite{Ferreira1} and for astrophysical measurements \cite{Ferreira2}. An updated analysis providing a discussion of the status quo is justified given the steady flow of new measurements (with gradually improving sensitivities), and also bearing in mind that this status quo is important for optimizing observational strategies of new astrophysical instruments such as ESPRESSO \cite{ESPRESSO} which in the coming years will significantly improve the sensitivity and reliability of these measurements. On the other hand, in our analysis we restrict ourselves to considering possible time variations of the fundamental couplings (corresponding to redshift dependencies in an astrophysical context). We will not consider the possibility of spatial (or environmental) dependencies; some constraints on such scenarios can be obtained using measurements of these constants within the Galaxy \cite{Galaxy}.

\section{Current local and spectroscopic measurements}

By repeatedly measuring and comparing the rates of two different atomic clocks one obtains a constraint on the relative shift of the corresponding characteristic frequencies. These will be proportional to certain products of fundamental couplings, and thus the measurement can be translated into a constraint of the drift of that combination. Different clock comparisons are sensitive to different products of these couplings, and therefore a combined analysis can in principle lead to constraints on each of them.

\begin{table*}
\begin{center}
\caption{Atomic clock constraints of varying fundamental couplings. The third, fourth and fifth columns show the sensitivity coefficients of each frequency ratio to the various dimensionless couplings.}
\label{table0} 
\begin{tabular}{|c|c|c c c|c|}
\hline
Clocks & ${\dot \nu_{AB}}/{\nu_{AB}}$ (yr${}^{-1}$) & $\lambda_\alpha$ & $\lambda_\mu$ & $\lambda_g$ & Reference \\ 
\hline
Hg-Al & $(5.3\pm7.9)\times10^{-17}$ & -2.95 & 0.0 &  0.000 & Rosenband {\it et al.} (2008) \protect\cite{Rosenband} \\
Dy162-Dy164 & $(-5.8\pm6.9)\times10^{-17}$ &  1.00 & 0.0 &  0.000 & Leefer {\it et al.} (2013) \protect\cite{Leefer} \\
Cs-SF${}_6$ & $(-1.9\pm2.7)\times10^{-14}$ &  2.83 & 0.5 & -1.266 & Shelkovnikov {\it et al.} (2008) \protect\cite{Shelkovnikov} \\
Cs-H & $(3.2\pm6.3)\times10^{-15}$ &  2.83 & 1.0 & -1.266 & Fischer {\it et al.} (2004) \protect\cite{Fischer} \\
Cs-Sr & $(1.80\pm0.55)\times10^{-16}$ &  2.77 & 1.0 & -1.266 & Abgrall {\it et al.} (2015) \protect\cite{Abgrall} \\
Cs-Hg & $(-3.7\pm3.9)\times10^{-16}$ &  5.77 & 1.0 & -1.266 & Fortier {\it et al.} (2007) \protect\cite{Fortier} \\
Cs-Yb(E2) & $(-0.5\pm1.9)\times10^{-16}$ &  1.83 & 1.0 & -1.266 & Tamm {\it et al.} (2014) \protect\cite{Tamm} \\
Cs-Yb(E3) & $(-0.2\pm4.1)\times10^{-16}$ &  8.83 & 1.0 & -1.266 & Huntemann {\it et al.} (2014) \protect\cite{Huntemann} \\
Cs-Rb & $(1.07\pm0.49)\times10^{-16}$ &  0.49 & 0.0 & -2.000 & Gu\'ena {\it et al.} (2012) \protect\cite{Bize2} \\
\hline
\end{tabular}
\end{center}
\end{table*}

Typically the ratio of two atomic clock frequencies will be proportional to
\be
\nu_{AB}=\frac{\nu_A}{\nu_B}\propto \alpha^{\lambda_\alpha} \mu^{\lambda_\mu} g_p^{\lambda_g}
\ee
where the $\lambda_i$ are the sensitivity coefficients
\be
\lambda_\alpha=\frac{d\ln{\nu_{AB}}}{d\ln\alpha}\,,
\ee
and analogously for the other couplings. A compilation of currently available measurements and their sensitivity coefficients can be found in Table \ref{table0} (reprinted from Table 6 of \cite{ROPP}); there are a total of nine different measurements (each relying on a comparison between different pairs of clocks). Some comparisons are only sensitive to $\alpha$, while others are sensitive to various products of $\alpha$, $\mu$ and $g_p$. Since we have nine measurements and only three unknowns (the individual values of the drifts of each of the three couplings) one can therefore use all the data to extract constraints on each of them.

\begin{figure*}
\begin{center}
\includegraphics[width=3.2in,keepaspectratio]{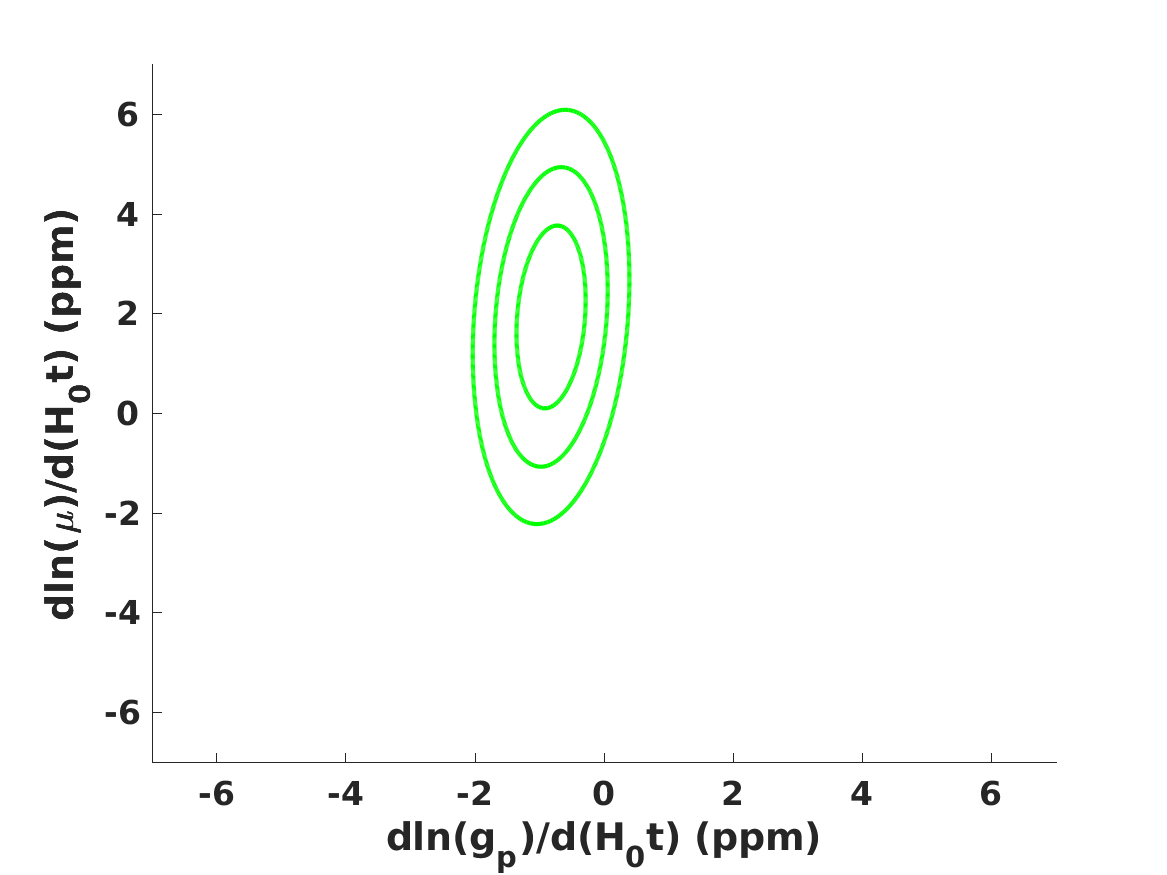}
\includegraphics[width=3.2in,keepaspectratio]{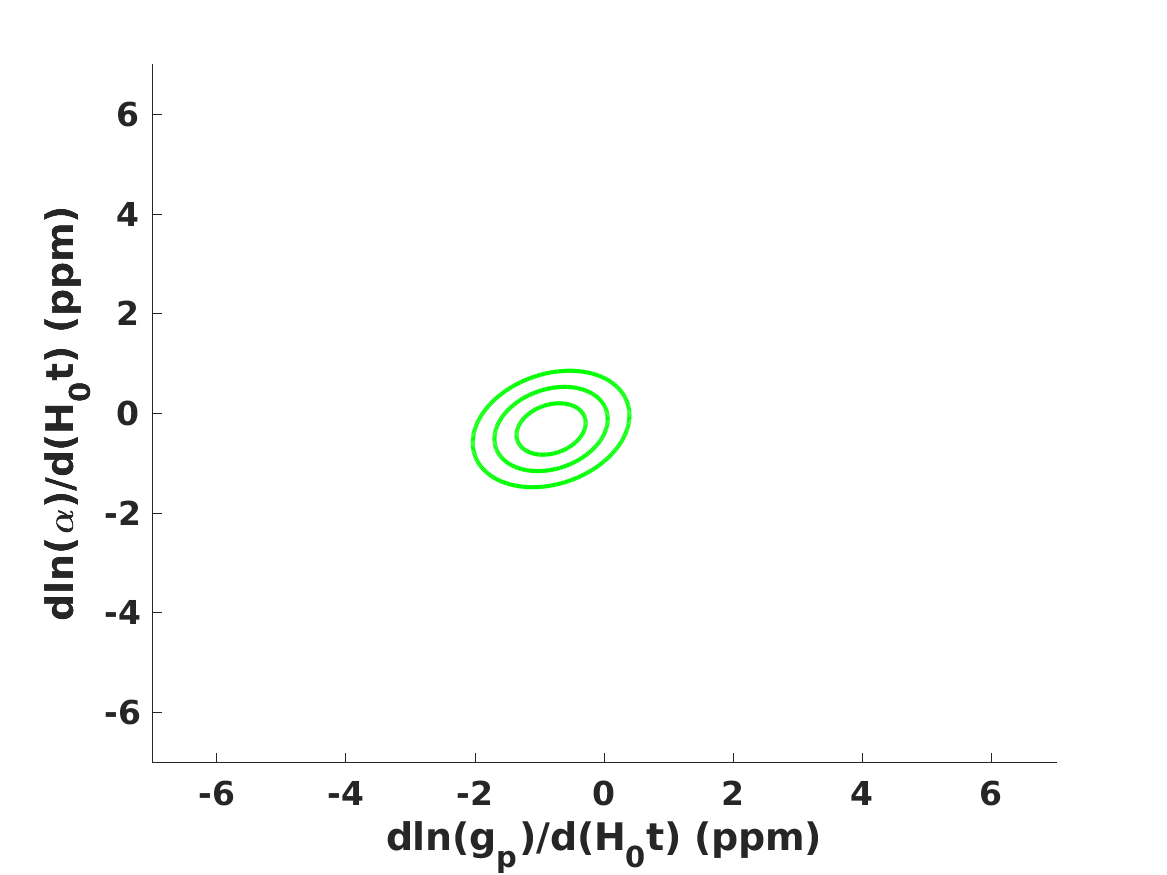}
\includegraphics[width=3.2in,keepaspectratio]{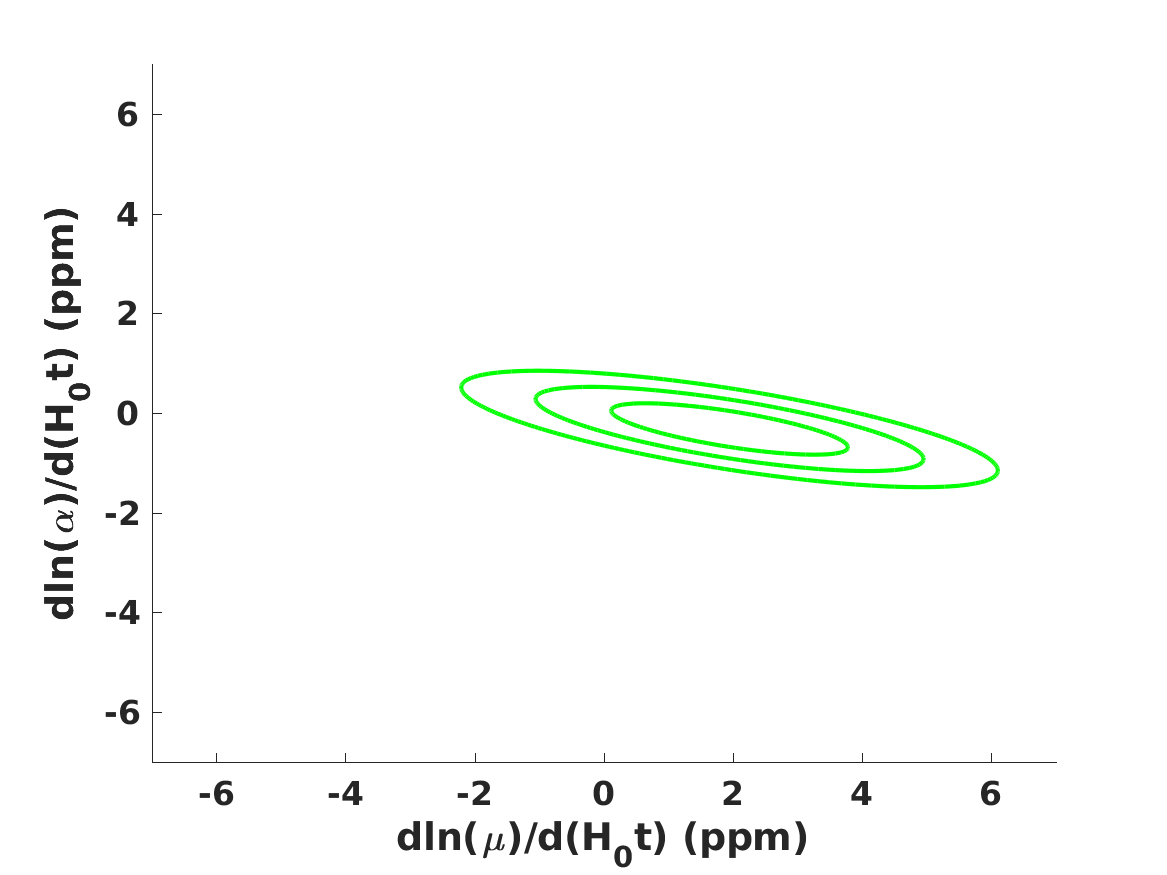}
\includegraphics[width=3.2in,keepaspectratio]{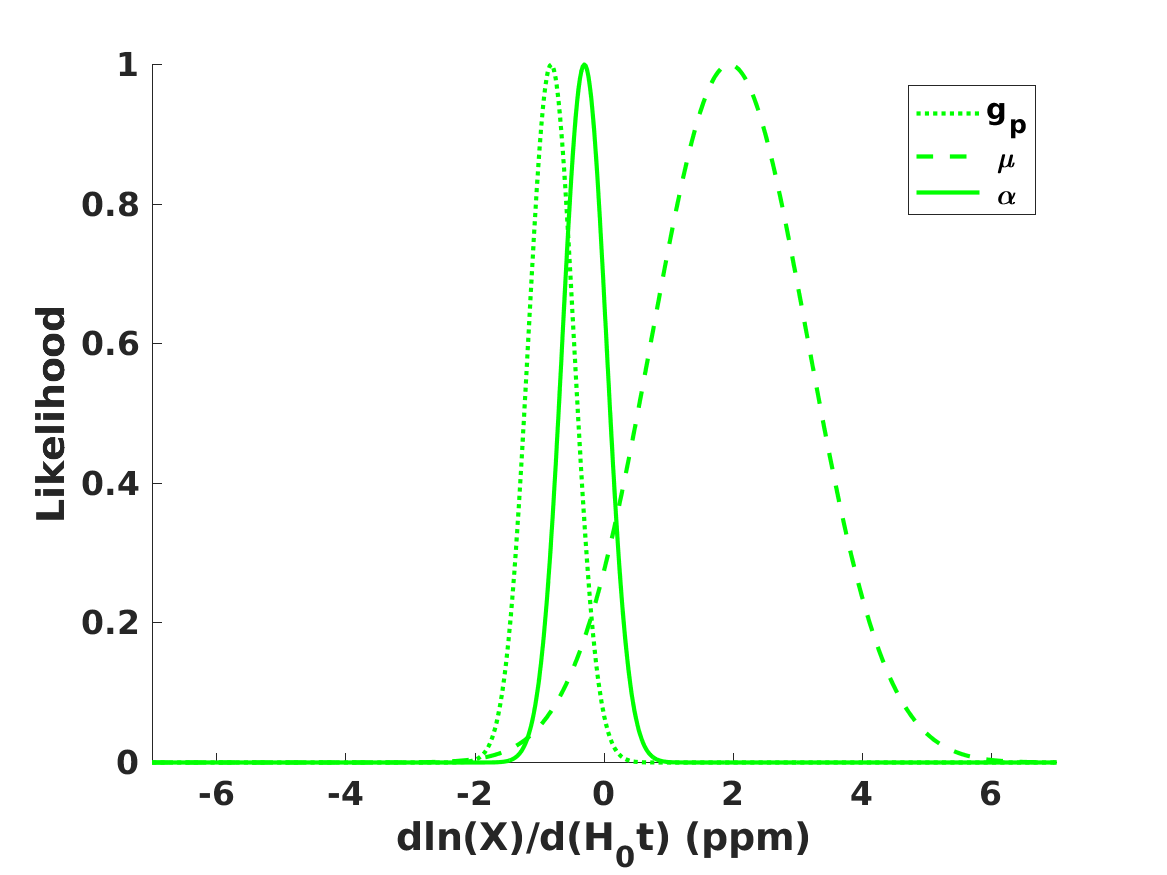}
\end{center}
\caption{\label{fig1}Atomic clock constraints on the current dimensionless drift rates of $\alpha$, $\mu$ and $g_p$, in ppm units. The first three panels show the one, two and three sigma confidence regions in the various two-dimensional planes (with the remaining parameter marginalized), while the final panel shows the posterior for each coupling.}
\end{figure*}

Figure \ref{fig1} shows the results of this analysis. Note that atomic clocks constrain the present-day drift rates of the relevant parameters, so these constraints are usually presented in units of inverse time. Here, for the purpose of a more intuitive comparison with astrophysical measurements (see below) we will present them in dimensionless form by dividing them by the Hubble parameter, which we assume to have the value $H_0=70$ km/s/Mpc. It is clear that changing this value by a few percent in either direction (as suggested by cosmological or low-redshift measurements respectively \cite{HPlanck,HRiess}) will not significantly impact our results. Moreover, we will typically present results in parts per million (henceforth ppm). Thus the current individual constraints from atomic clocks are
\be
\frac{d\ln{\alpha}}{H_0dt}=(-0.31\pm0.34)\, ppm\,,\quad (\Delta\chi^2=0.83)
\ee
\be
\frac{d\ln{\mu}}{H_0dt}=(1.9\pm1.2)\, ppm\,,\quad (\Delta\chi^2=2.6)
\ee
\be
\frac{d\ln{g_p}}{H_0dt}=(-0.82\pm0.36)\, ppm\,,\quad (\Delta\chi^2=5.5)\,.
\ee
In each case we also provide (to two significant digits) the $\Delta\chi^2$ between the null result and the best-fit value. We see that there is no evidence for a drift in $\alpha$; this is consistent with the result of atomic clock measurements which are only sensitive to $\alpha$, the more stringent of which is \cite{Rosenband}. Nevertheless, the global analysis provides very mild evidence for drifts in $g_p$ and $\mu$, with opposite signs. Note also that the absolute error on the drift of $\mu$ is larger than one ppm, while those on the other two constants are significantly below ppm level.  This difference is due to the fact that the three strongest constraints in Table \ref{table0} are insensitive to $\mu$.

While atomic clock comparisons measure drift rates at redshift $z=0$, high resolution astrophysical spectroscopy can measure the values of the couplings (or combinations thereof) at specific redshifts. It is possible to measure $\alpha$ and $\mu$ individually, while so far $g_p$ has only been measured in various products with one or both of the other couplings.

For $\alpha$, there is a large dataset by Webb {\it et al.} \cite{Webb}, containing a total of 293 archival measurements from the HIRES and UVES spectrographs, respectively at the Keck and VLT telescopes, which has been described extensively in the literature. In an attempt to verify these results (considering its possible vulnerability to systematics in both spectrographs \cite{Syst}), several groups have been collecting dedicated measurements, for which the data collection and analysis attempts to minimize this systematics. A compilation of 21 such measurements can be found in Table 1 of \cite{ROPP}, and additional measurements have been published in \cite{Cooksey}. For $\mu$, a compilation of 16 recent measurements can be found in Table 2 of \cite{ROPP}. As for the measurements of products of $\alpha$, $\mu$ and $g_p$, Table 3 of \cite{ROPP} lists the 29 measurements then available, and more recently two stringent ones have been published in \cite{Kanekar,Gupta}. All of these will constitute our astrophysical measurements dataset.

\begin{figure*}
\begin{center}
\includegraphics[width=3.2in,keepaspectratio]{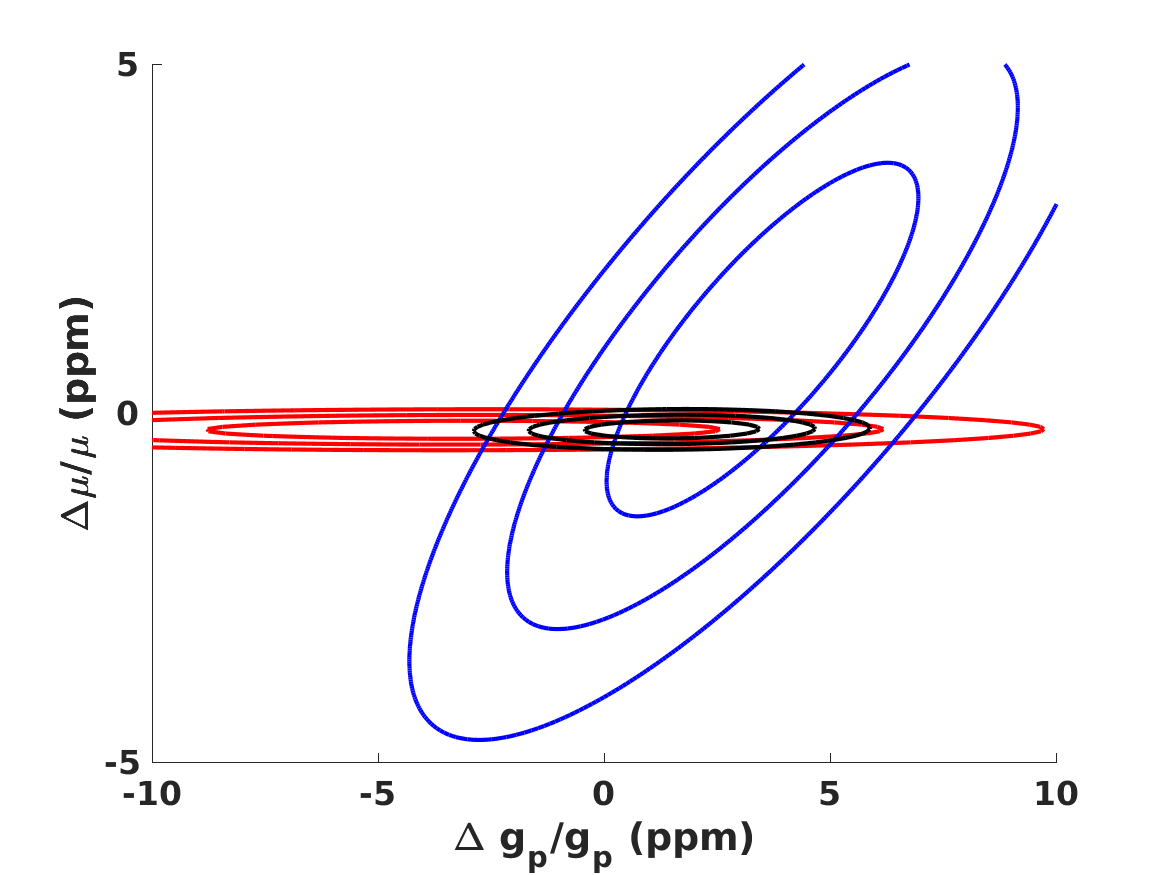}
\includegraphics[width=3.2in,keepaspectratio]{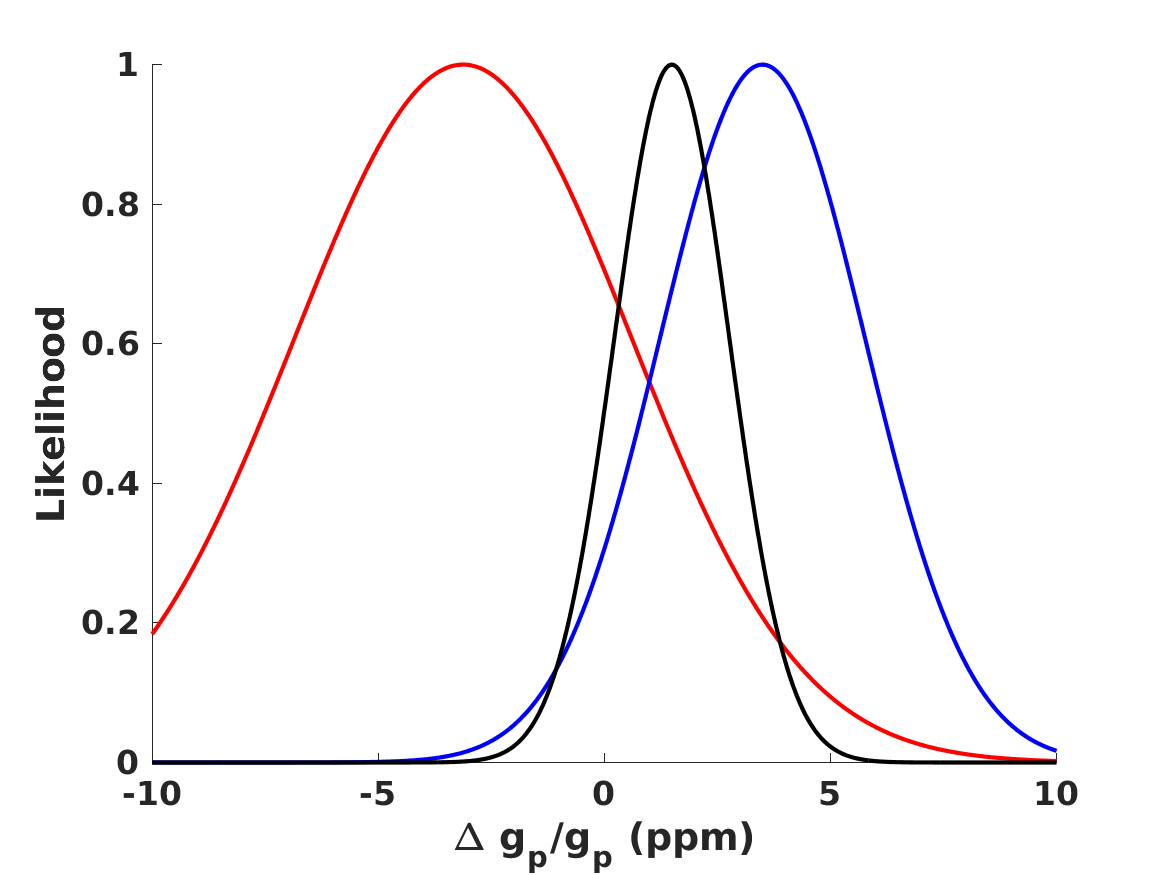}
\includegraphics[width=3.2in,keepaspectratio]{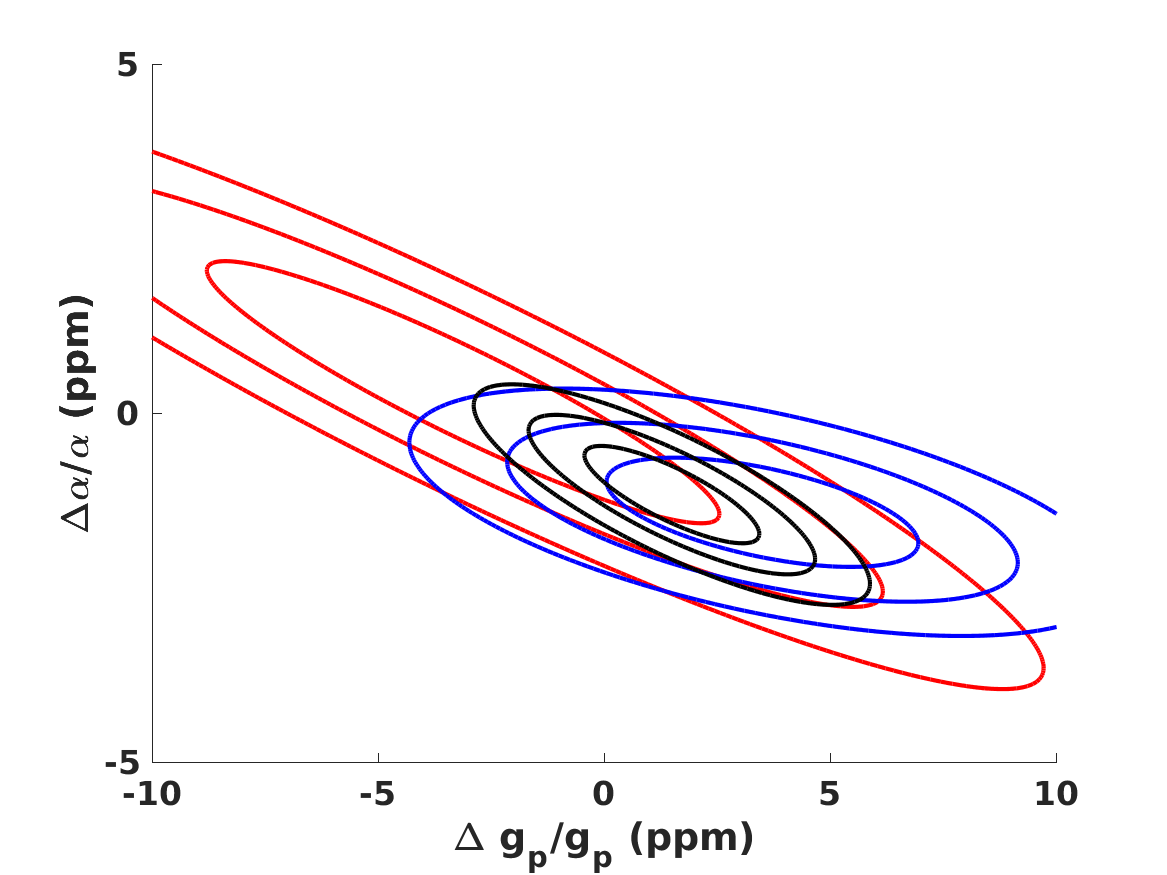}
\includegraphics[width=3.2in,keepaspectratio]{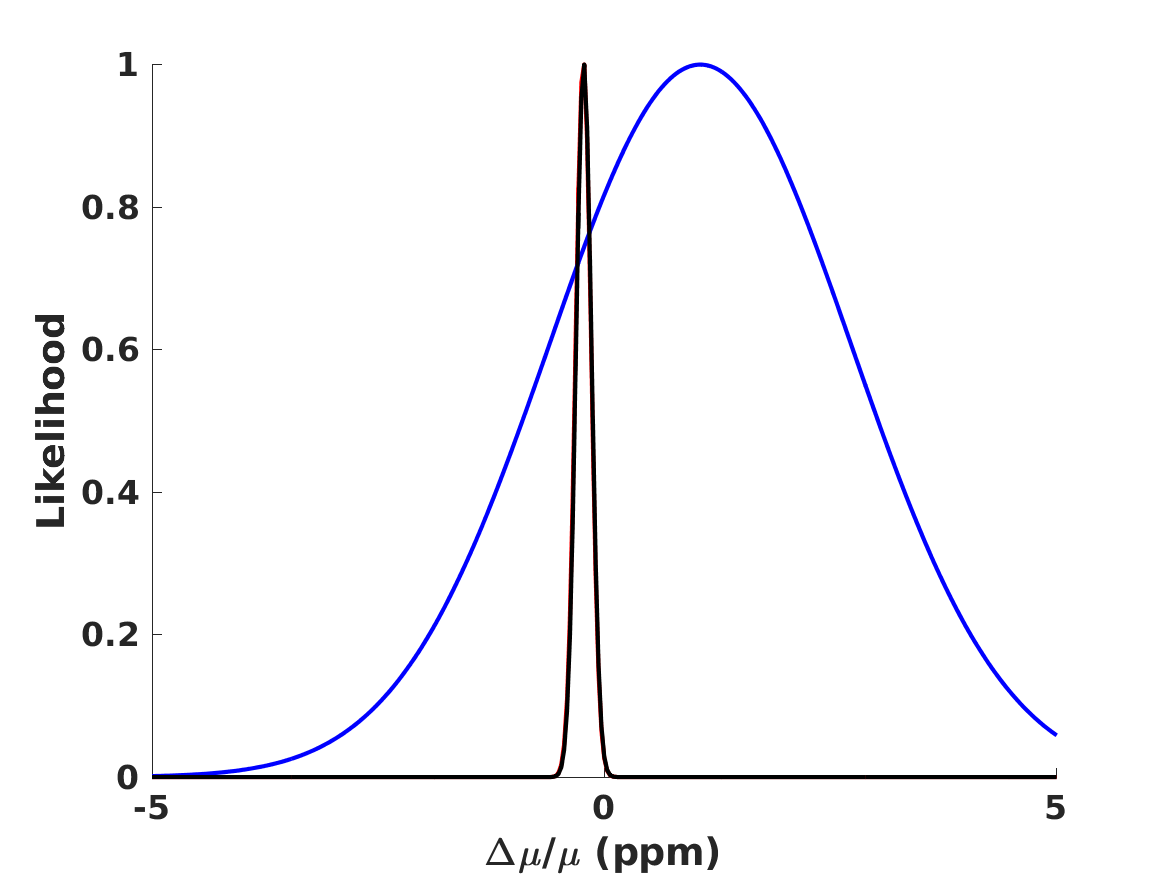}
\includegraphics[width=3.2in,keepaspectratio]{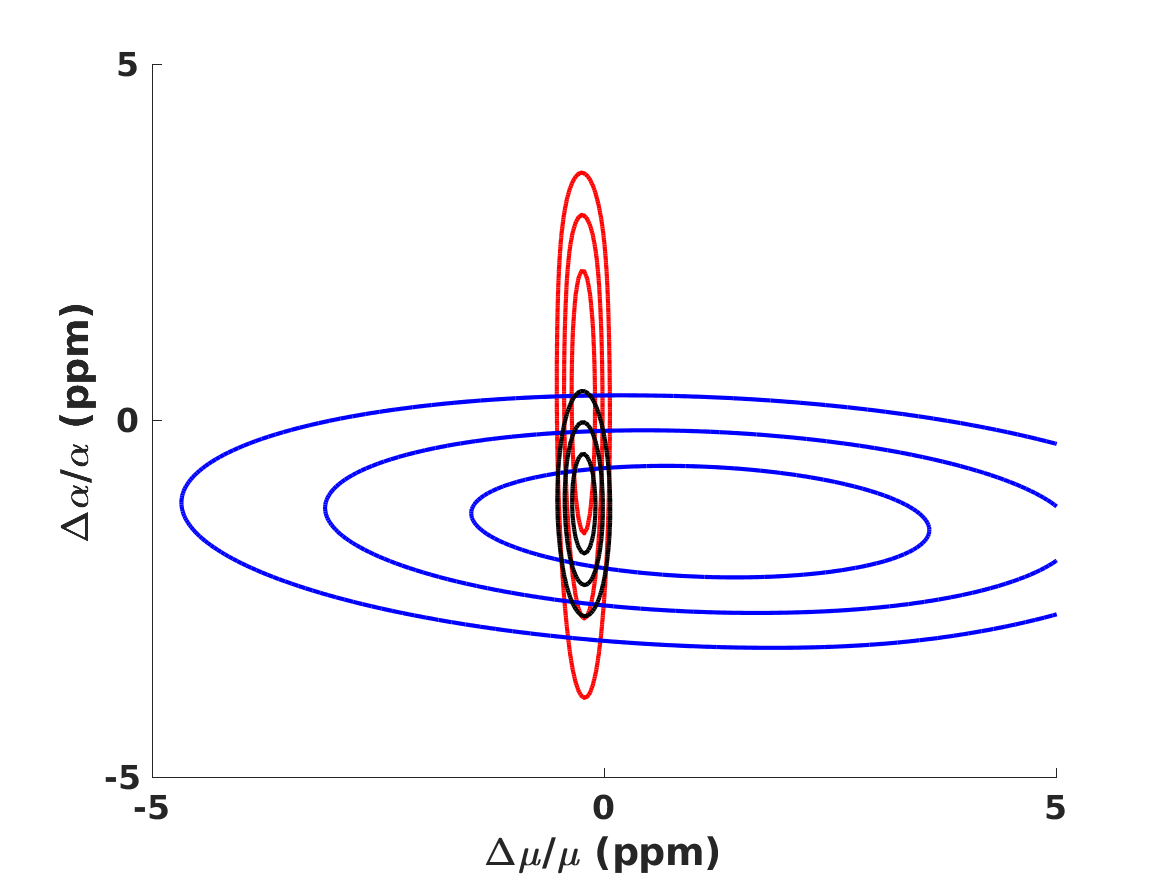}
\includegraphics[width=3.2in,keepaspectratio]{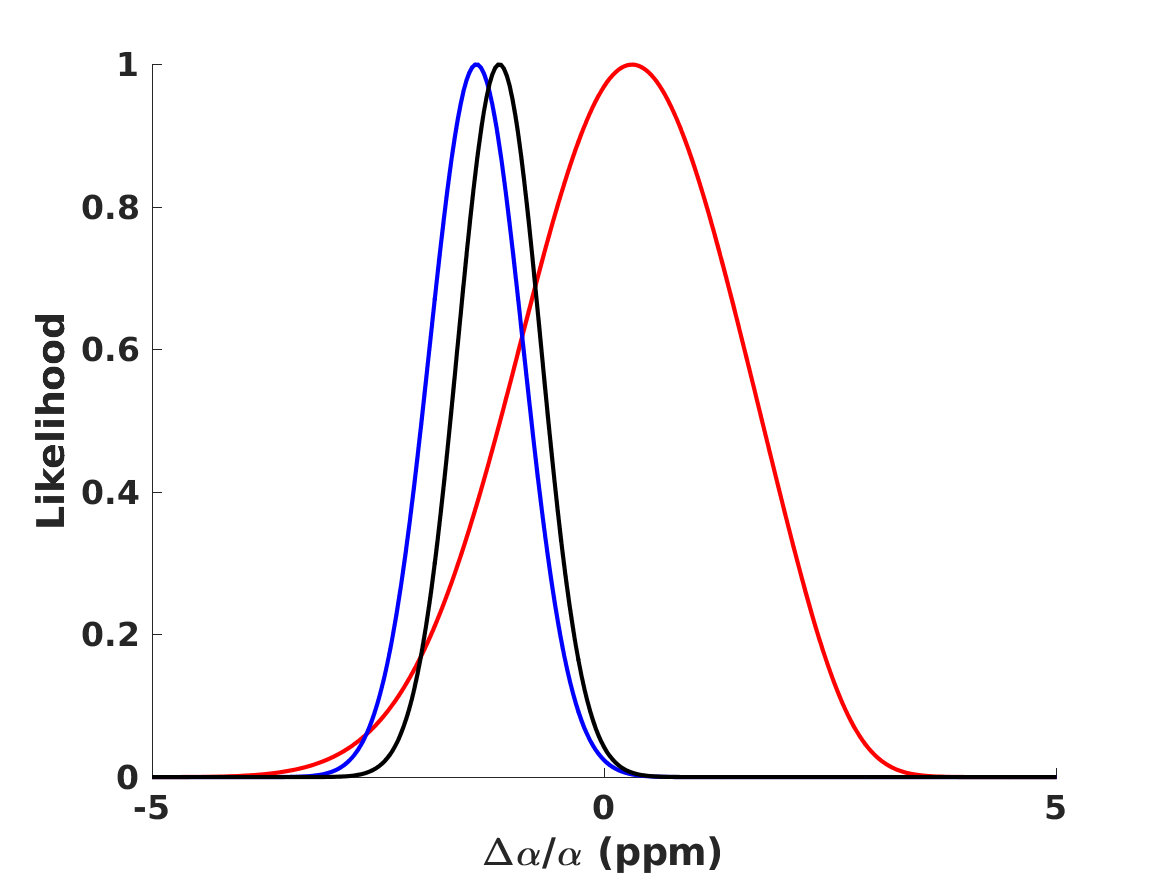}
\end{center}
\caption{\label{fig2}Astrophysical constraints on relative variations of $\alpha$, $\mu$ and $g_p$ with respect to their laboratory values, in ppm units. The left panels show the one, two and three sigma confidence regions in the various two-dimensional planes (with the remaining parameter marginalized), while the right panels show the posterior for each coupling. In all cases the red curves correspond to low redshift data ($z<1$), the blue curves to high redshift data ($z>1$) and the black curves to the full dataset.}
\end{figure*}

A joint analysis can now be done as in the case of atomic clocks, though there are two differences. The first is that astrophysical observations measure the values of these couplings at different redshifts, rather than their drift rates. We will generally express them as relative variations with respect to their local (laboratory) value. For example for $\alpha$ we define
\be
\frac{\Delta\alpha}{\alpha} (z)=\frac{\alpha(z)-\alpha_0}{\alpha_0}\,,
\ee
with $\alpha_0$ being the laboratory value, and similarly for the other couplings. Note that we do expect these quantities to be redshift dependent. Since different measurements are done at different redshifts, and currently span a redshift range from $z\sim0$ to beyond $z=6$, we will need to make a choice regarding the binning of the data. Since it is clear that the current data only allows meaningful results in a limited number of bins, we will split our dataset in only two bins, $z<1$ and $z>1$, as well as considering the full dataset. This choice is in part motivated by theoretical considerations (with the low redshift bin corresponding to the epoch of acceleration and the high redshift bin to the matter era), but it is also well justified from an observational perspective. An illustration of this point is the fact that all the available direct measurements of $\mu$ from radio/mm transitions have been done at redshifts $z<1$ (there are no known high redshift targets for which the required transitions have been observed) while all measurements of $\mu$ from optical transitions have been done at redshifts $z>2$ (at lower redshift the required transitions fall in the ultraviolet and can only be observed from space).

\begin{table*}
\begin{center}
\caption{One sigma ($68.3\%$ confidence level) constraints on the three fundamental couplings (with the others marginalized), for the low and high redshift data as well as for all redshifts combined. We show separately the constraints from the complete datasets and from the dedicated measurements only (i.e., the latter do not include the archival measurements of $\alpha$). All constraints are in ppm units. In each case we also provide (to two significant digits) the $\Delta\chi^2$ between the null result and the best-fit value.}
\label{table1}
\begin{tabular}{| l | c c | c c | c c |}
\hline
{ } & \multicolumn{2}{c|}{$z<1$} & \multicolumn{2}{c|}{$z>1$} & \multicolumn{2}{c|}{All redshifts} \\
\hline
Parameter & Constraint & $\Delta\chi^2$ & Constraint & $\Delta\chi^2$ & Constraint & $\Delta\chi^2$ \\
\hline
$\frac{\Delta\alpha}{\alpha}$ (All data) & $0.32\pm1.27$ & $0.06$ & $-1.43\pm0.52$ & $7.5$ & $-1.18\pm0.46$ & $6.4$ \\
$\frac{\Delta\alpha}{\alpha}$ (Dedicated only) & $1.62\pm1.51$ & $1.1$ & $-0.98\pm0.62$ & $2.7$ & $-0.73\pm0.55$ & $1.8$ \\
\hline
$\frac{\Delta\mu}{\mu}$(All data) & $-0.23\pm0.09$ & $7.4$ & $1.05\pm1.68$ & $0.40$ & $-0.23\pm0.09$ & $7.1$ \\
$\frac{\Delta\mu}{\mu}$(Dedicated only) & $-0.22\pm0.10$ & $7.5$ & $0.86\pm1.68$ & $0.26$ & $-0.22\pm0.10$ & $7.3$ \\
\hline
$\frac{\Delta g_p}{g_p}$ (All data) & $-3.1\pm3.7$ & $0.70$ & $3.5\pm2.3$ & $2.4$ & $1.5\pm1.2$ & $1.4$ \\
$\frac{\Delta g_p}{g_p}$ (Dedicated only) & $-7.3\pm5.1$ & $2.1$ & $2.5\pm2.4$ & $1.1$ & $0.5\pm1.4$ & $0.15$ \\
\hline
\end{tabular}
\end{center}
\end{table*}

With these caveats, the results of this analysis are summarized in Figure \ref{fig2} and also Table \ref{table1}. Note that in the former we show only the constraints from the complete datasets, while in the latter we show both the constraints from the complete datasets and those from the recent dedicated measurements (which excludes the archival measurements of $\alpha$). One sees that there is mild (two sigma) evidence for the variation of $\mu$ at low redshifts, and for the variation of $\alpha$ at high redshift (although in the latter case this becomes only one sigma if the Webb {\it et al.} date is not included). For $g_p$, we find consistency with the null result at the two sigma level in all cases.

Having done separate global analyses for both the local and astrophysical measurements, one may naturally ask how they relate to each other. A qualitative comparison can easily be done. For example, in the case of $\mu$, the low-redshift data prefers a smaller value of $\mu$ in the past, which is consistent with the atomic clocks' preference for a positive drift rate. However, for a more quantitative comparison one needs to specify an evolution model. We will address this in the following sections.

\section{Relating local and astrophysical data}

In what follows we will assume that the physical mechanism underlying the evolution of the constants is the dynamics of a scalar field. For simplicity we will assume that this is a simple dilaton-type field \cite{Campbell}. One can show that under these assumptions the evolution of $\alpha$ can be described by a single model parameter \cite{Damour,Runaway}
\be
\frac{\Delta\alpha}{\alpha}=\kappa \ln{(1+z)}\,.
\ee
One can easily calculate the time derivative of this expression, which allows us to express the drift rate of $\alpha$ as a function of the free parameter $\kappa$,
\be
\kappa=-\left(\frac{1}{H}\frac{d\ln{\alpha}}{dt}\right)_0\,.
\ee
Thus in this simplest model the local drift and cosmological variation are related without any free parameters
\be
\frac{\Delta\alpha}{\alpha}=-\left(\frac{1}{H}\frac{d\ln{\alpha}}{dt}\right)_0 \ln{(1+z)}\,. \label{Dilz}
\ee
We will assume that analogous relations hold for $\mu$ and $g_p$; this is a valid assumption for a wide class of models, as we will discuss in some detail in the following section. This allows us to compare the local and astrophysical measurements.

\begin{figure*}
\begin{center}
\includegraphics[width=3.2in,keepaspectratio]{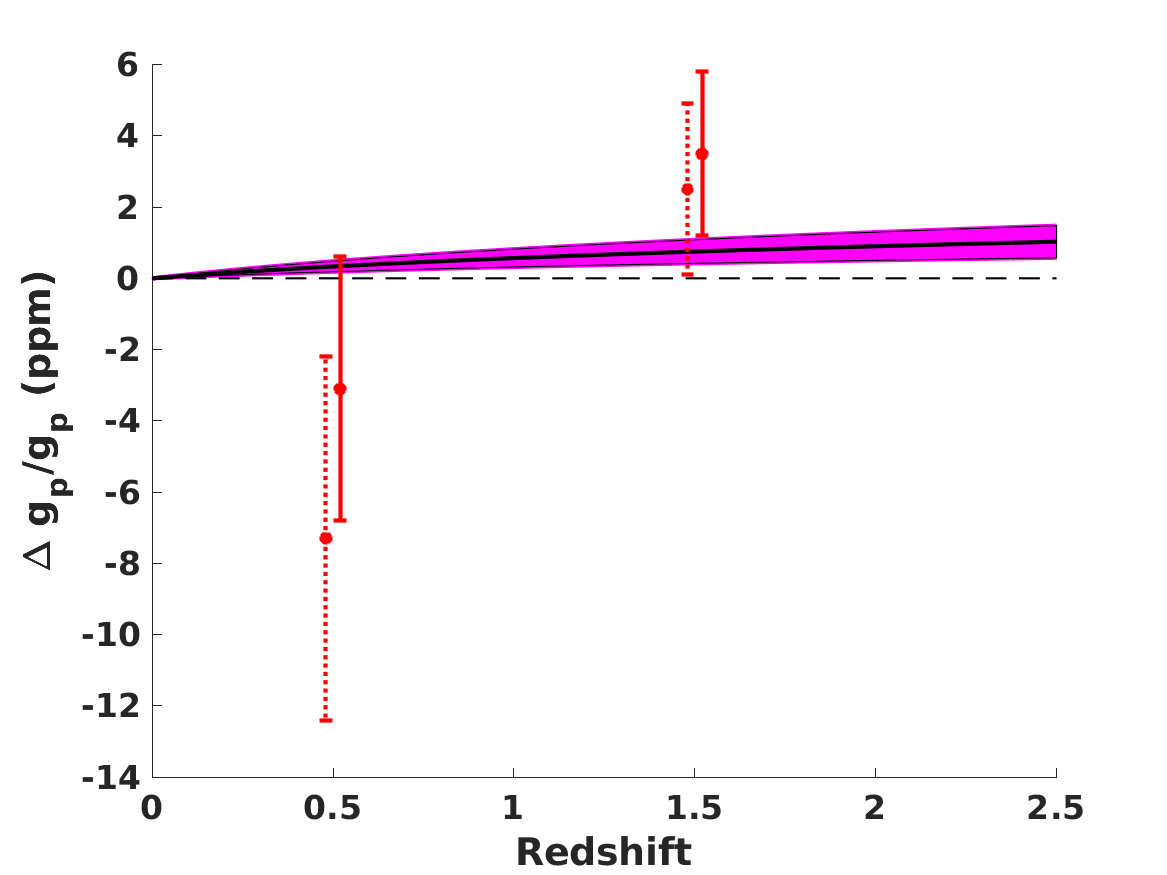}
\includegraphics[width=3.2in,keepaspectratio]{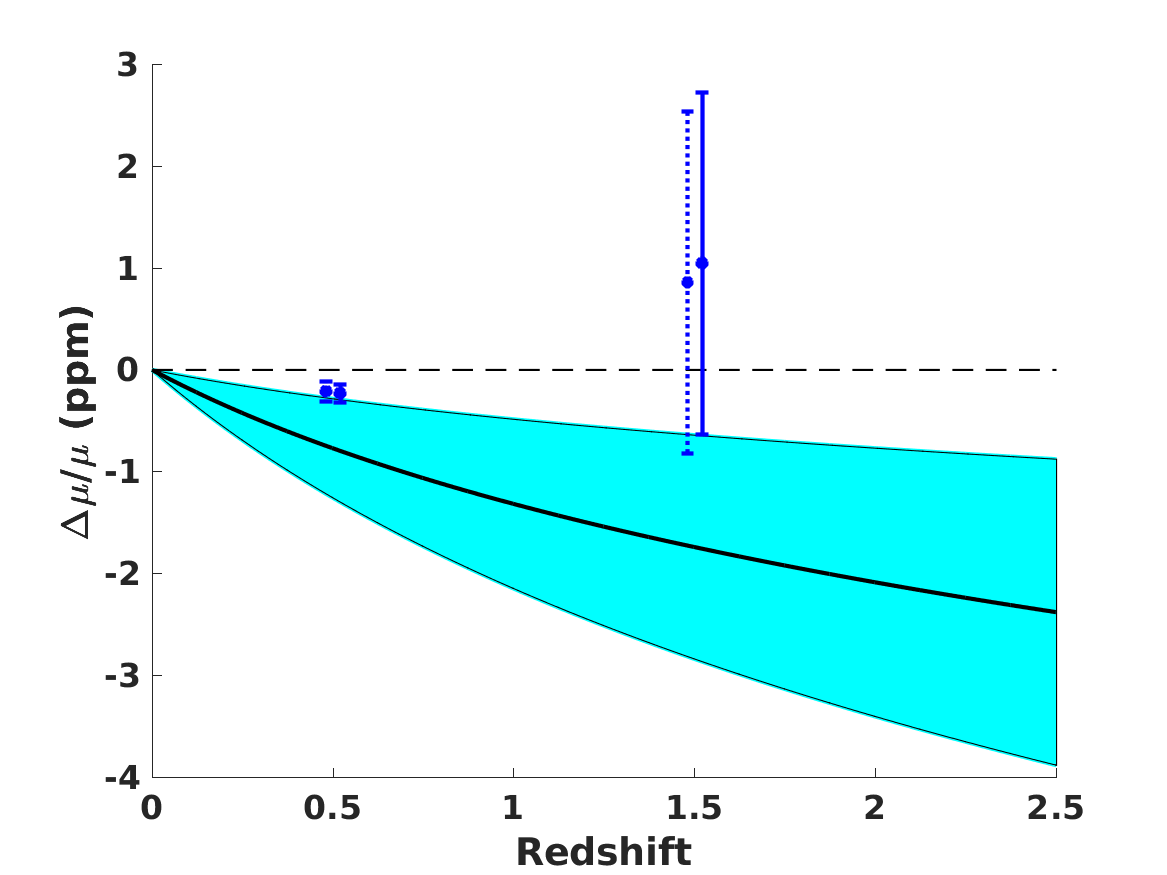}
\includegraphics[width=3.2in,keepaspectratio]{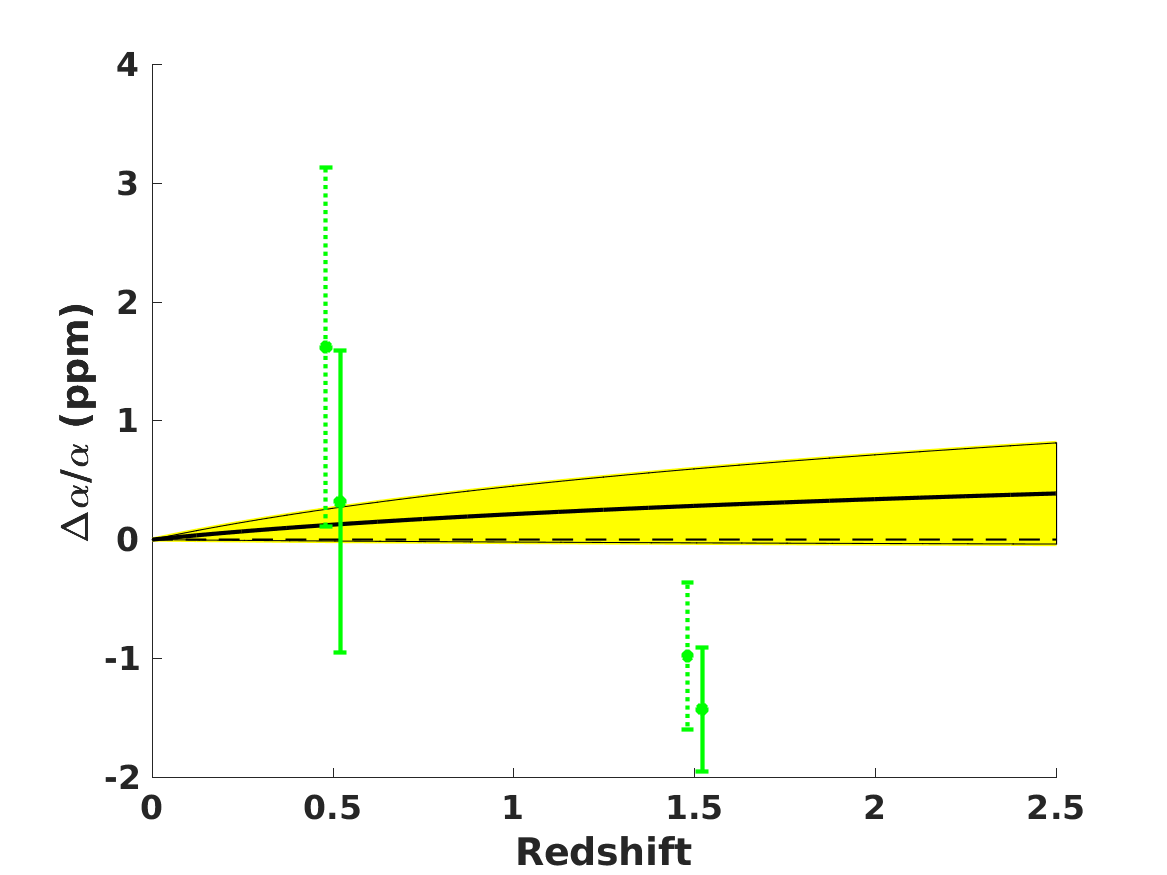}
\end{center}
\caption{\label{fig3}Comparing local and astrophysical constraints on $\alpha$, $\mu$ and $g_p$. The colored bands depict the range of values of each constant compatible (at the one sigma confidence level) with the local atomic clock constraint, assuming the evolution model of Eq. \protect\ref{Dilz}. The data points are the low redshift and high redshift one-sigma constraints of Table \protect\ref{table1} (with the results for the dedicated measurements depicted with the dotted error bars and those for the full datasets by the solid error bars), which have been nominally placed at around redshifts $z=0.5$ and $z=1.5$ respectively.}
\end{figure*}

Figure \ref{fig3} shows the results of this comparison. The colored bands show the range of values of each coupling which are consistent at each redshift (and at the one sigma level) with the atomic clocks constraints. The low and high redshift constraints in each case have also been depicted in each panel, both for the dedicated measurements alone and for the complete datasets. We have nominally placed these constraints around the redshifts $z=0.5$ and $z=1.5$, but these represent combined constraints from a range of different redshifts.

Overall we find the low redshift data to be in reasonable agreement with the atomic clock constraints, except for $g_p$ when one only uses the dedicated measurements. On the other hand, the high redshift data is somewhat discrepant, more so in the the case where the archival data of Webb {\it et al.} is included: there is disagreement at one sigma, but they do agree at two sigma. These plots also make it clear that astrophysical constraints are typically less stringent than the atomic clock ones (at least when compared in this way), but the low redshift measurement of $\mu$ is an exception to this.

\section{Relating variations in different couplings}

In Section 2 we have assumed that the variations of $\alpha$, $\mu$ and $g_p$ are independent: given say a relative variation of $\alpha$ we have not specified whether the corresponding variations of $\mu$ and $g_p$ are larger or smaller, by how much, or even whether or not the variations have the same or opposite signs. In realistic particle physics models the variations of the three constants will not be independent, but the relations between them will depend on the specific model chosen. In what follows we study some representative examples of these models.

For this we follow \cite{Coc,Luo}, considering a class of grand unification models (in which unification happens at some unspecified high energy) in which the weak scale is determined by dimensional transmutation and further assuming that relative variations of all the Yukawa couplings are the same. We maintain the assumption (already introduced in the previous section)  that the variation of the couplings is driven by a dilaton-type scalar field (as in \cite{Campbell}). With these assumptions one finds that the variations of $\mu$ and $\alpha$ are related through
\begin{equation}
\frac{\Delta\mu}{\mu}=[0.8R-0.3(1+S)]\frac{\Delta\alpha}{\alpha}\,,
\end{equation}
where $R$ and $S$ can be taken as free phenomenological (model-dependent) parameters. They are defined as follows
\begin{equation}
\frac{\Delta \Lambda_{QCD}}{\Lambda_{QCD}} = R \frac{\Delta \alpha}{\alpha}
\end{equation}
\begin{equation}
\frac{\Delta \nu}{\nu} = S \frac{\Delta h}{h}
\end{equation}
where $\Lambda_{QCD}$ is the QCD mass scale, $\nu$ is the vacuum expectation value of the Higgs and $h$ stands for any of the Yukawa couplings (we're assuming that all of them vary in the same way). Their absolute value can be anything from order unity to several hundreds, although physically one usually expects them to be positive. Similarly, for the proton gyromagnetic ratio one finds \cite{flambaum1,flambaum2}
\begin{equation}
\frac{\Delta g_p}{g_p}=[0.10R-0.04(1+S)]\frac{\Delta\alpha}{\alpha}\,.
\end{equation}

These relations allow us to relate any measurement of $\mu$ or $g_p$ combination of constants (for example $\alpha$, $\mu$ and $g_p$) into a measurement of $\alpha$, although only in a model-dependent way. Note that in this class of models the proton gyromagnetic ratio is less sensitive to the parameters $R$ and $S$ than the proton-to-electron mass ratio.

In what follows we will consider three specific examples of these models, each described by specific values of the unification parameters $R$ and $S$. As discussed by \cite{Coc,Langacker}, current (possibly naive) expectations regarding unification scenarios suggest that typical values are
\begin{equation}
R\sim36\,,\quad S\sim160\,;
\end{equation}
we refer to this as the Unification model. Although these numbers may be representative, they are certainly not unique. As an example we take the dilaton-type model whose variations of fundamental couplings have been discussed by \cite{Nakashima}, which find
\begin{equation}
R\sim109.4\,,\quad S\sim0\,;
\end{equation}
we refer to this as the Dilaton model. Finally, we take a third case
\begin{equation}
R=0\,,\quad S=-1\,;
\end{equation}
this phenomenologically corresponds to the case where $\alpha$ is allowed to vary while $\mu$ and $g_p$ are not. We will refer to this as the Bekenstein model, since one class of models with such a behaviour has been proposed in \cite{Bekenstein}.

\begin{figure*}
\begin{center}
\includegraphics[width=3.2in,keepaspectratio]{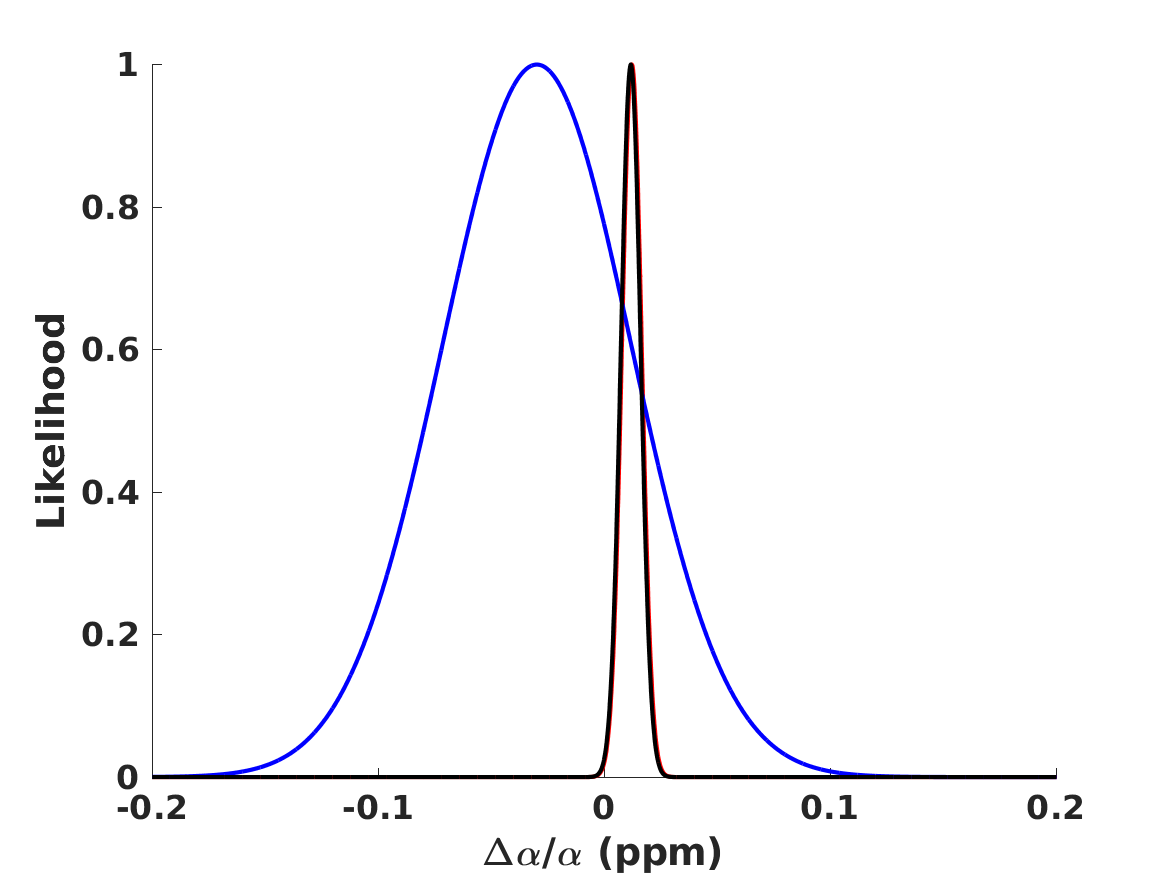}
\includegraphics[width=3.2in,keepaspectratio]{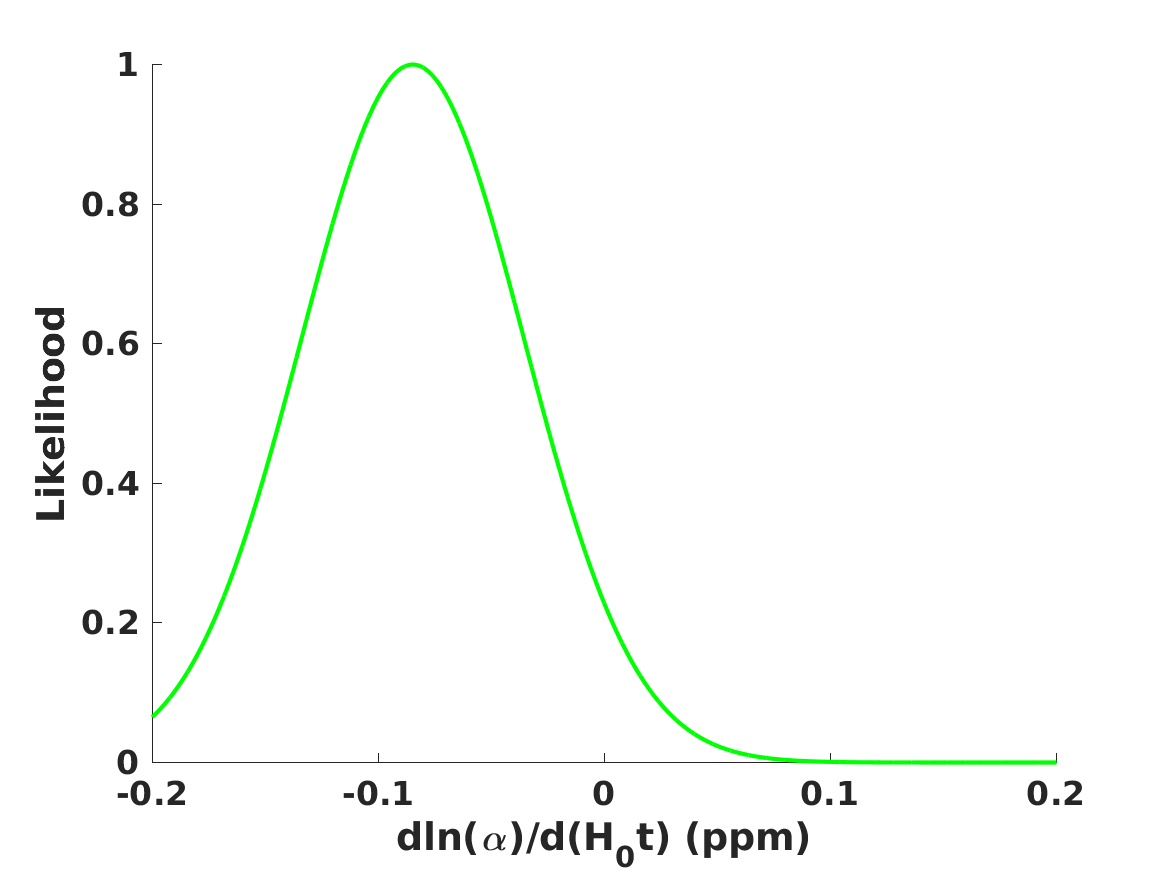}
\includegraphics[width=3.2in,keepaspectratio]{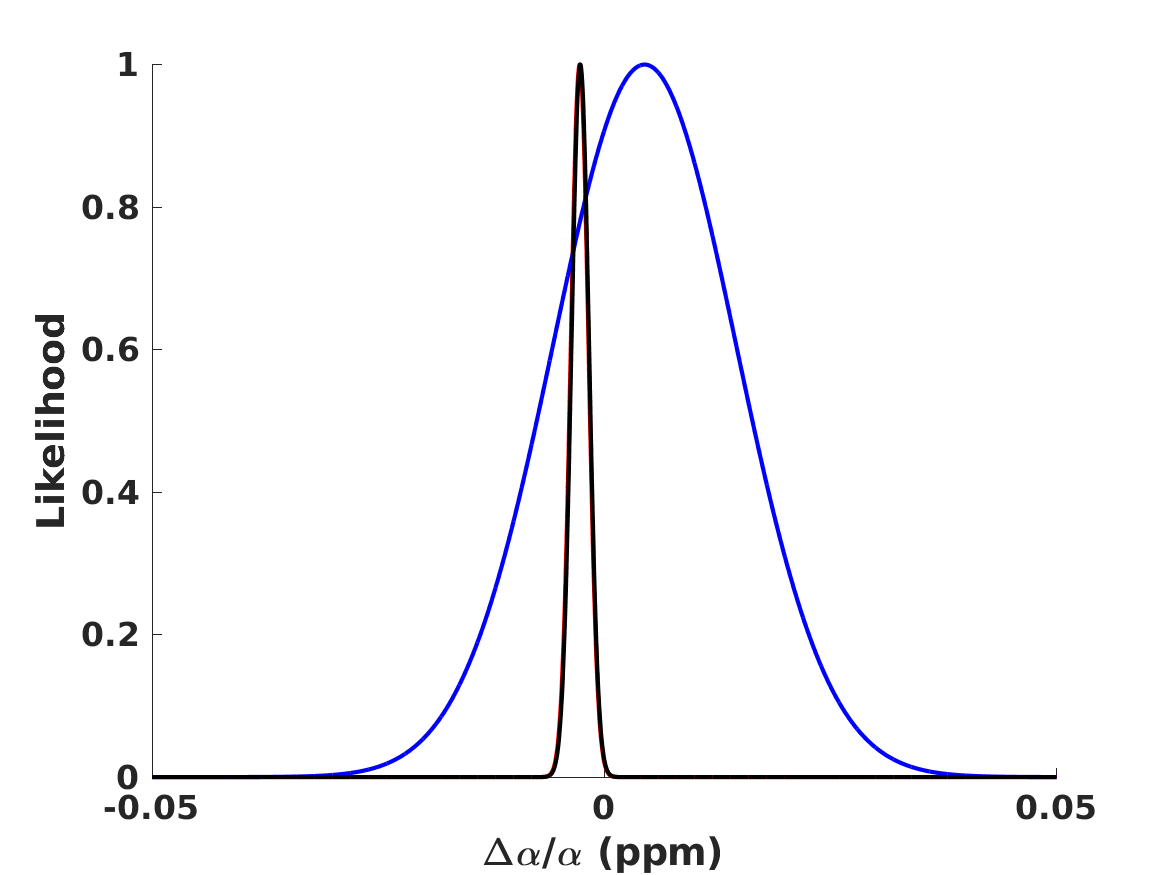}
\includegraphics[width=3.2in,keepaspectratio]{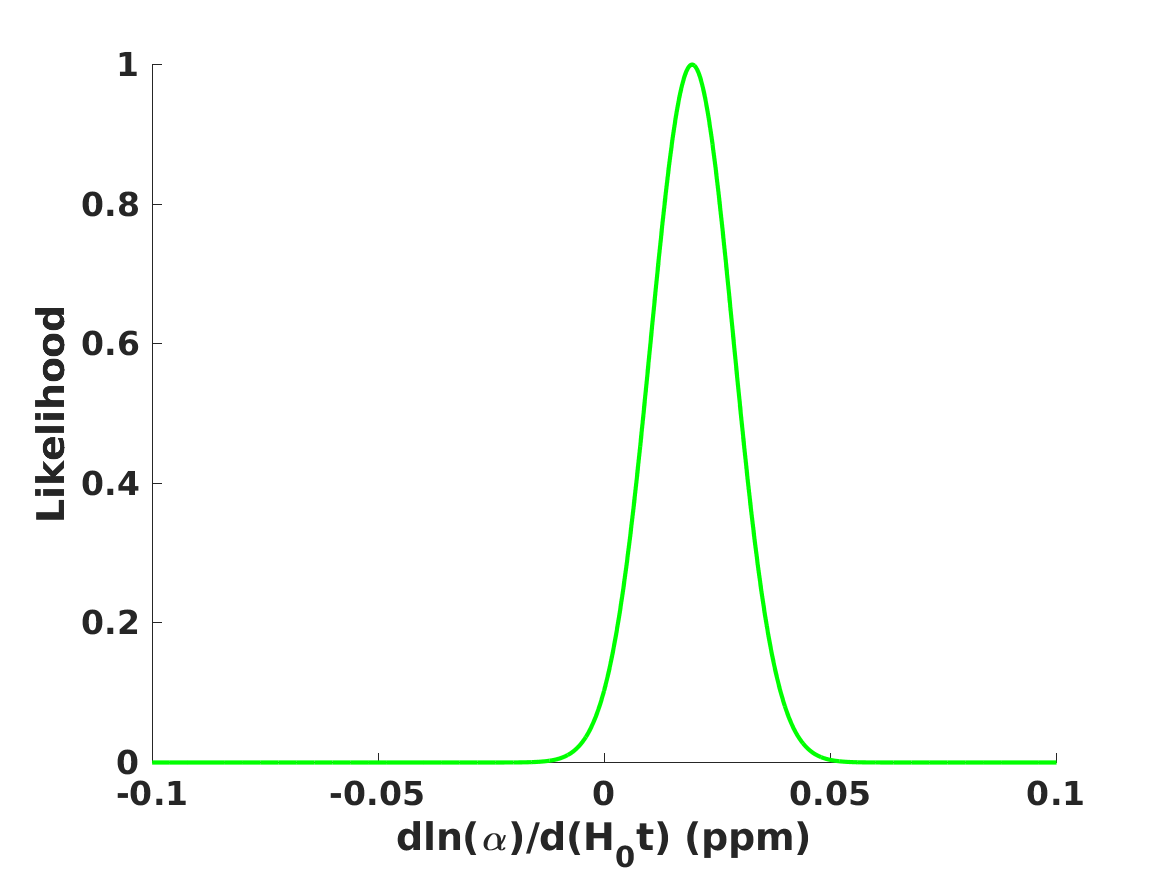}
\includegraphics[width=3.2in,keepaspectratio]{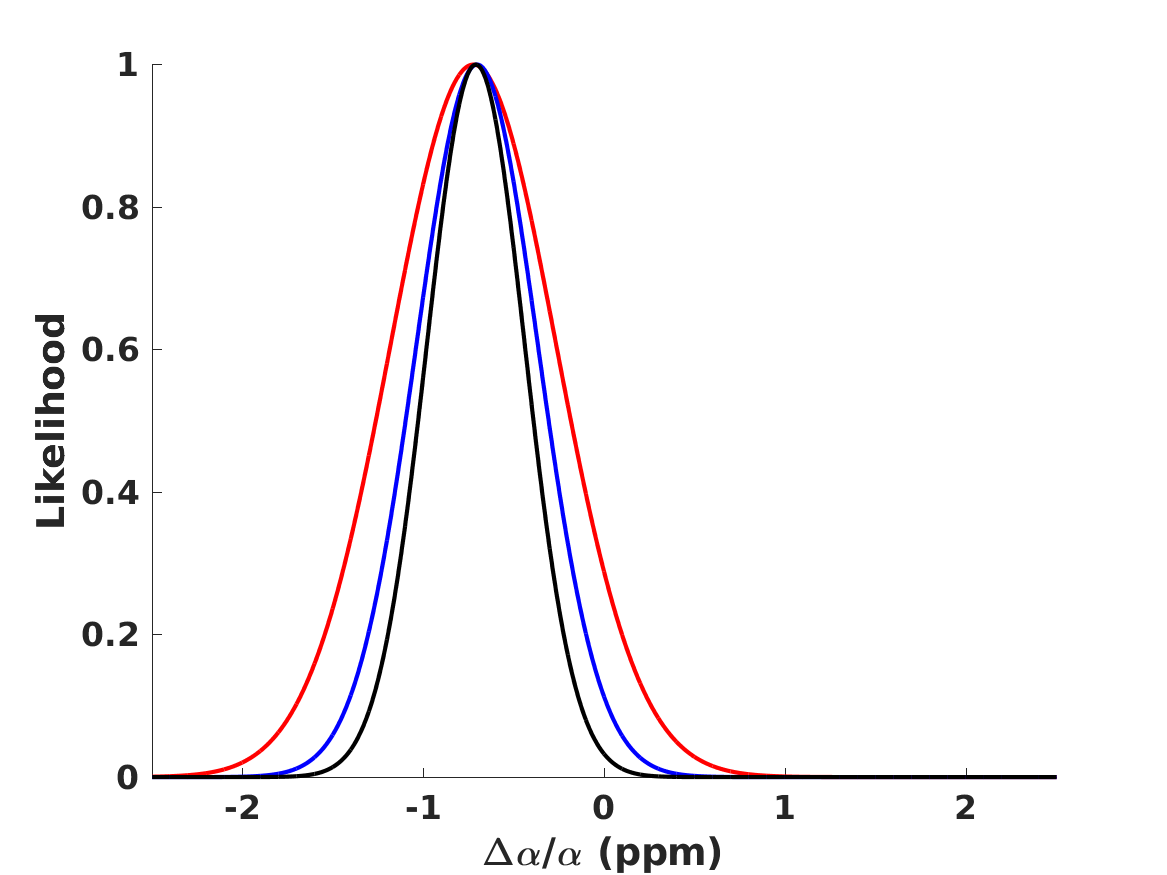}
\includegraphics[width=3.2in,keepaspectratio]{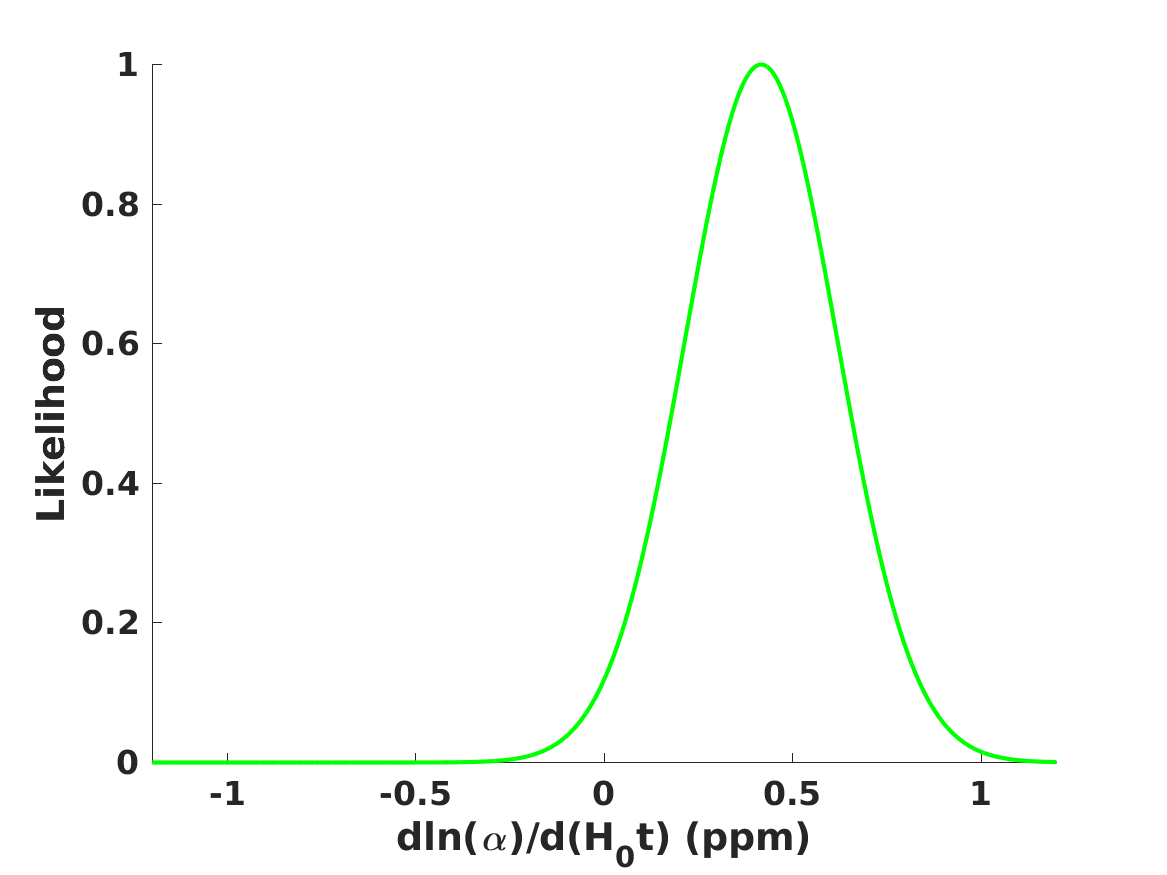}
\end{center}
\caption{\label{fig4}Astrophysical constraints on the relative variations of $\alpha$ with respect to their laboratory values (left panels), and atomic clock constraints on its current drift (right panels), in ppm units. The top, middle and bottom panels correspond respectively to the Unification, Dilaton and Bekenstein models. In the left panels the red curves correspond to low redshift data ($z<1$), the blue curves to high redshift data ($z>1$) and the black curves to the full dataset.}
\end{figure*}

For each model these relations are expected to be universal, so they can be applied both to astrophysical and to atomic clock measurements. We can thus repeat our earlier analysis for the three models; the results of this are displayed in Figure \ref{fig4} an in Table \ref{table2}; in the latter we again show the results for the full datasets and for the recent dedicated measurements only (which do not include the $\alpha$ archival data).

\begin{table*}
\begin{center}
\caption{One sigma ($68.3\%$ confidence level) constraints on the relative variation of $\alpha$ for the low and high redshift data as well as for all redshifts combined, and on its current drift, in the three models being considered.  We show separately the constraints from the complete datasets and from the dedicated measurements only (i.e., the latter do not include the archival measurements of $\alpha$). All constraints are in ppm units. In each case we also provide (to two significant digits) the $\Delta\chi^2$ between the null result and the best-fit value.}
\label{table2}
\begin{tabular}{| l | l | c c | c c | c c |}
\hline
{ } & { } & \multicolumn{2}{c|}{Unification} & \multicolumn{2}{c|}{Dilaton} & \multicolumn{2}{c|}{Bekenstein} \\
\hline
Data & Parameter & Constraint & $\Delta\chi^2$ & Constraint & $\Delta\chi^2$ & Constraint & $\Delta\chi^2$ \\
\hline
Atomic clocks & $\frac{d\ln{\alpha}}{d(H_0t)}$ & $(-8.5\pm4.9)\times10^{-2}$ & $3.0$ & $(1.9\pm0.9)\times10^{-2}$ & $4.5$ & $(4.2\pm2.0)\times10^{-1}$ & $4.2$ \\
\hline
{ } & $\frac{\Delta\alpha}{\alpha}$ ($z<1$) & $(1.2\pm0.4)\times10^{-2}$ & $7.7$ & $(-2.7\pm1.1)\times10^{-3}$ & $7.8$ & $(-7.2\pm4.6)\times10^{-1}$ & $2.5$ \\
QSO (All) & $\frac{\Delta\alpha}{\alpha}$ ($z>1$) & $(-3.0\pm4.2)\times10^{-2}$ & $0.51$ & $(4.4\pm10.2)\times10^{-3}$ & $0.19$ & $(-7.0\pm3.4)\times10^{-1}$ & $4.4$ \\
{ } & $\frac{\Delta\alpha}{\alpha}$ (All z) & $(1.2\pm0.4)\times10^{-2}$ & $7.2$ & $(-2.7\pm1.0)\times10^{-3}$ & $7.5$ & $(-7.1\pm2.7)\times10^{-1}$ & $6.9$ \\
\hline
{ } & $\frac{\Delta\alpha}{\alpha}$ ($z<1$) & $(1.2\pm0.4)\times10^{-2}$ & $7.7$ & $(-2.7\pm1.1)\times10^{-3}$ & $7.8$ & $(-6.9\pm4.7)\times10^{-1}$ & $2.1$ \\
QSO (Dedicated) & $\frac{\Delta\alpha}{\alpha}$ ($z>1$) & $(-2.5\pm4.2)\times10^{-2}$ & $0.37$ & $(4.7\pm10.1)\times10^{-3}$ & $0.22$ & $(-4.6\pm3.6)\times10^{-1}$ & $1.7$ \\
{ } & $\frac{\Delta\alpha}{\alpha}$ (All z) & $(1.2\pm0.4)\times10^{-2}$ & $7.3$ & $(-2.7\pm1.0)\times10^{-3}$ & $7.5$ & $(-5.4\pm2.9)\times10^{-1}$ & $3.6$ \\
\hline
\end{tabular}
\end{center}
\end{table*}

Several remarks are immediately warranted. Firstly, note that the constraints for the different models can differ both in sign and in order of magnitude. In all models there is some mild evidence for a variation of $\alpha$ at low redshifts (at up to the three sigma level), while at high redshifts this is only the case for the Bekenstein model (and in this case the preferred value is approximately the same in both redshift bins, 0.7 ppm).

\begin{figure*}
\begin{center}
\includegraphics[width=3.2in,keepaspectratio]{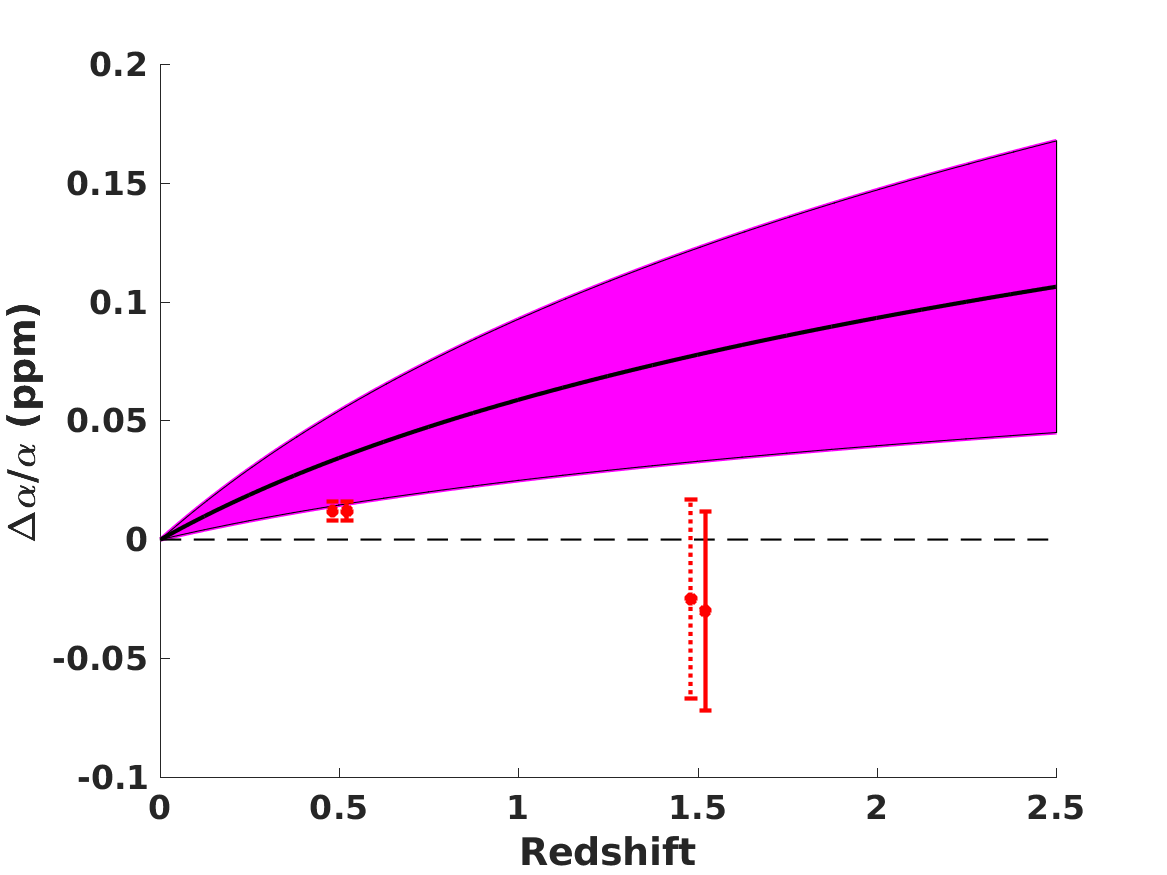}
\includegraphics[width=3.2in,keepaspectratio]{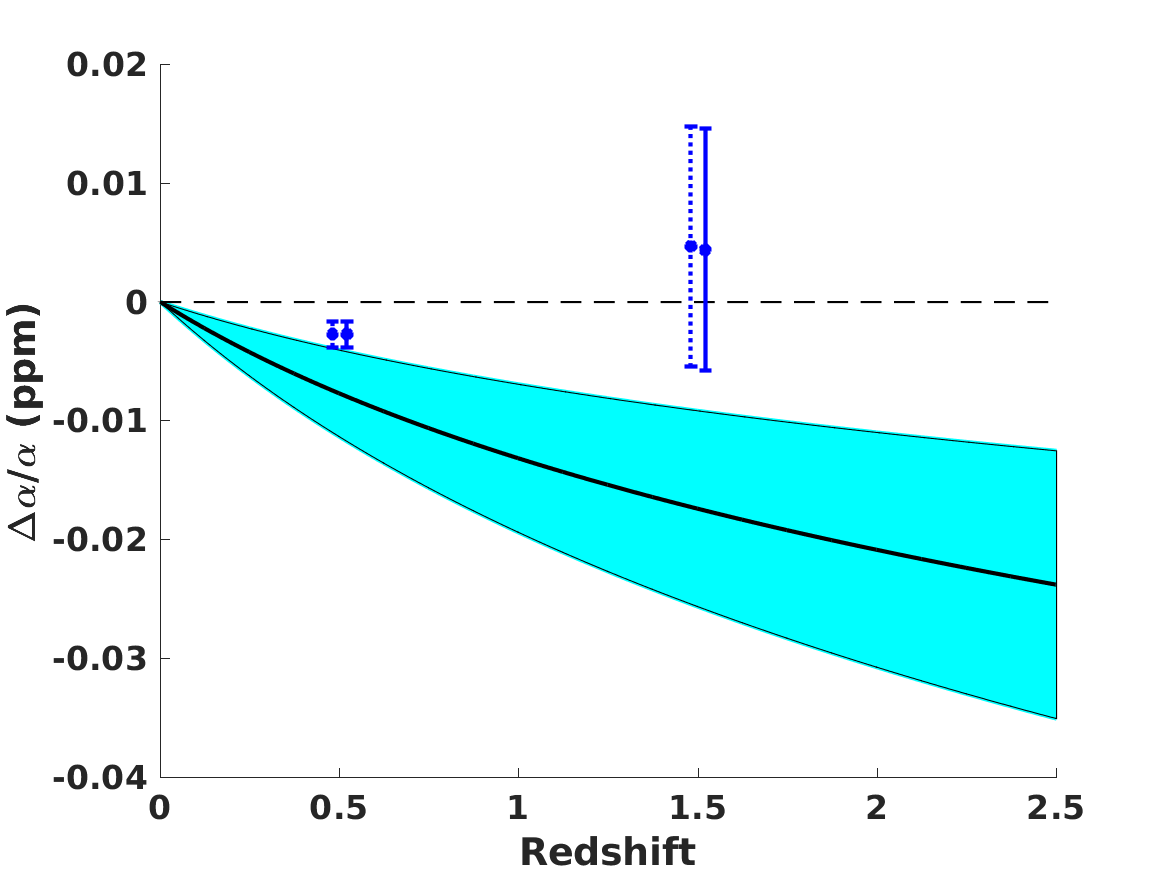}
\includegraphics[width=3.2in,keepaspectratio]{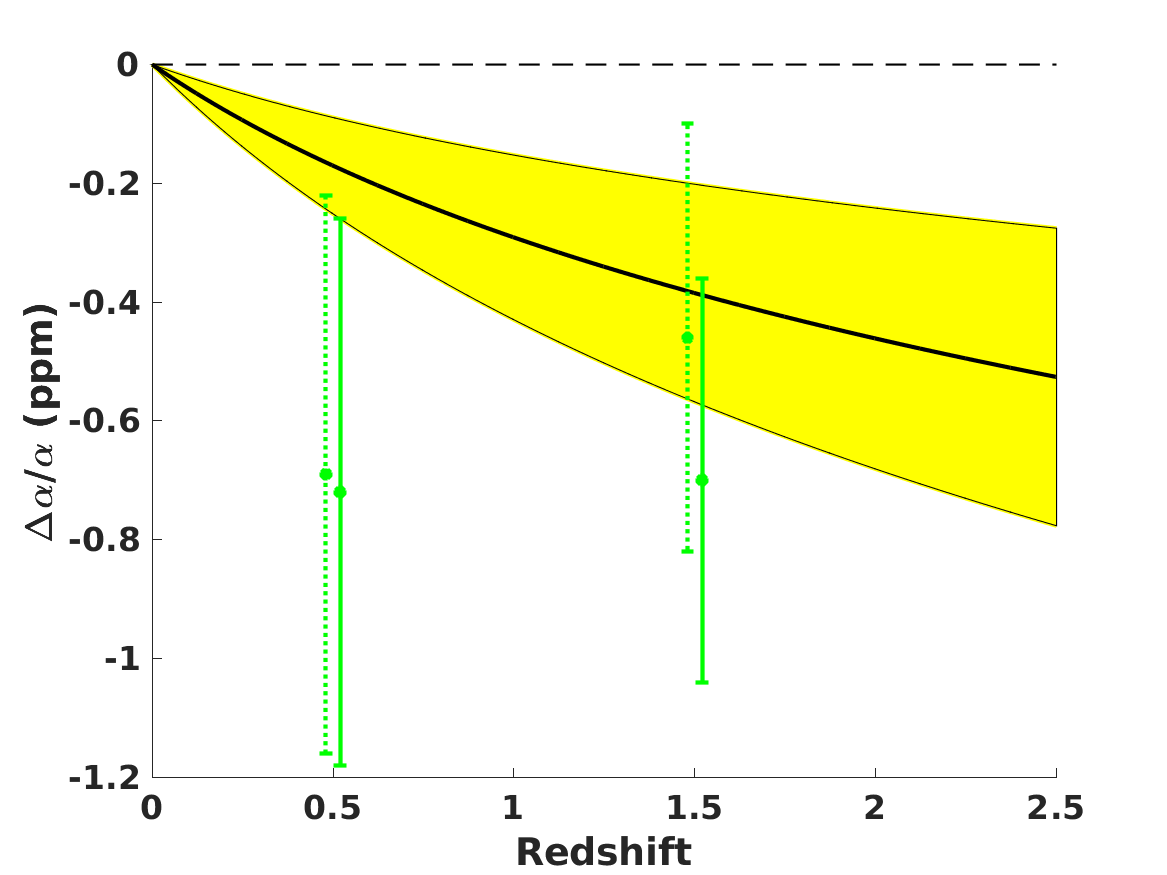}
\end{center}
\caption{\label{fig5}Comparing local and astrophysical constraints on $\alpha$ in the Unification, Dilaton and Bekenstein models. The colored bands depict the range of values of each constant compatible (at the one sigma confidence level) with the local atomic clock constraint, assuming the evolution model of Eq. \protect\ref{Dilz}. The data points are the low redshift and high redshift one-sigma constraints of Table \protect\ref{table2} (with the results for the dedicated measurements depicted with the dotted error bars and those for the full datasets by the solid error bars), which have been nominally placed around redshifts $z=0.5$ and $z=1.5$ respectively.}
\end{figure*}

It is also worthy of note that in all cases the current drift rates are positive when the preferred low redshift value is negative and vice-versa, so at least qualitatively these are in agreement. A more quantitative comparison can be seen in Figure \ref{fig5}. It is clear that there is agreement between the atomic clock and the low redshift data for all three models, but the high redshift data is only in agreement with the atomic clock data for the Bekenstein model. Lastly, for this purpose there is no substantial difference between using the dedicated measurements only or the full datasets, although the overall agreement is slightly better in the former case.

\begin{figure*}
\begin{center}
\includegraphics[width=3.2in,keepaspectratio]{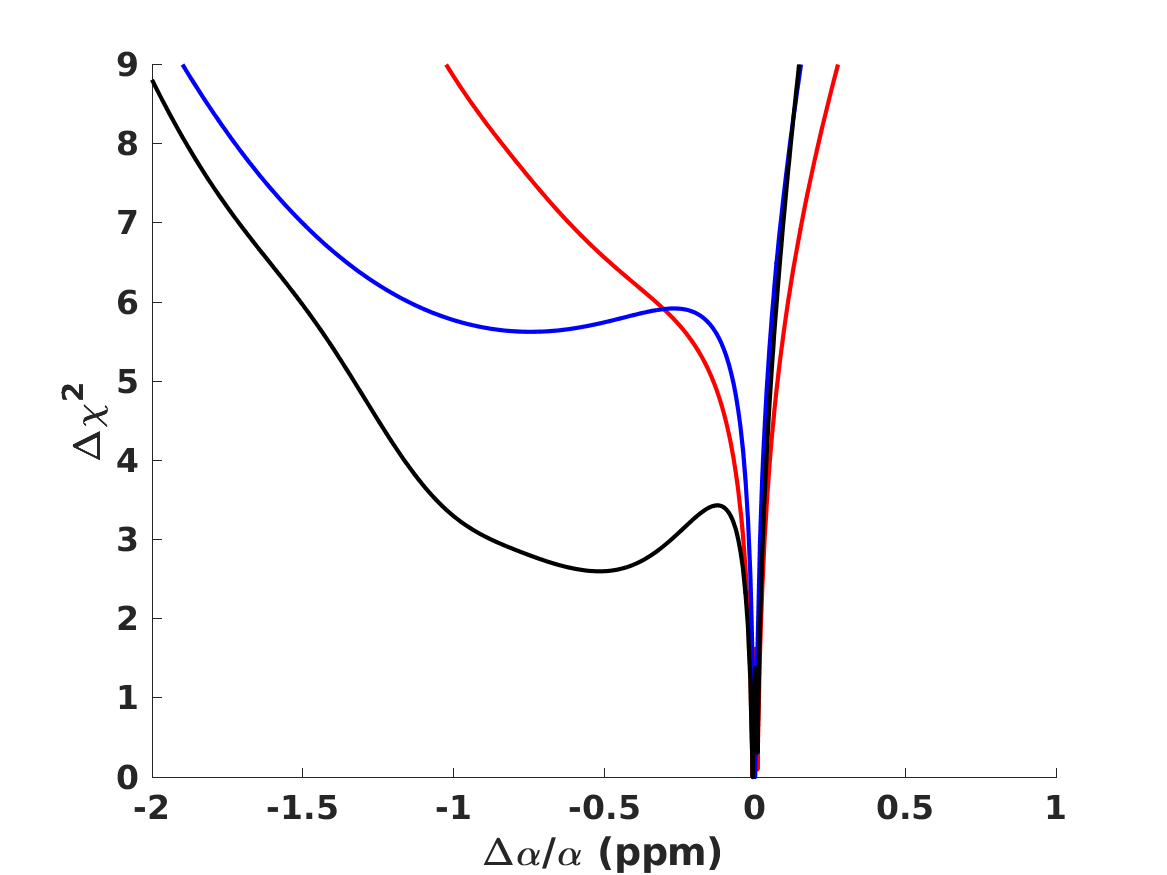}
\includegraphics[width=3.2in,keepaspectratio]{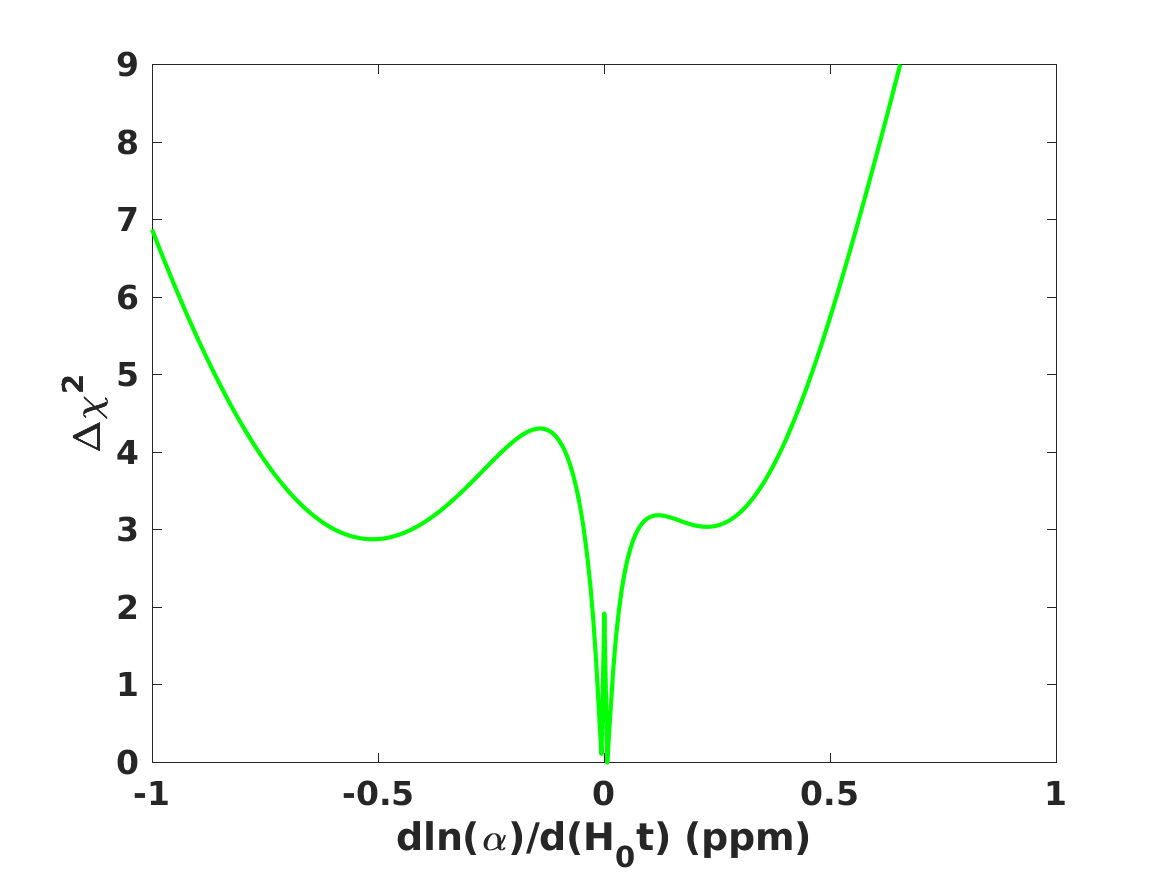}
\end{center}
\caption{\label{fig6}Astrophysical constraints on the relative variations of $\alpha$ with respect to their laboratory values (left panel), and atomic clock constraints on its current drift (right panel), in ppm units, with the $R$ and $S$ parameters marginalized. Both panels show the chi-square relative to its minimum value, $\Delta\chi^2=\chi^2-\chi^2_{min}$. In the left panel the red curve corresponds to low redshift data ($z<1$), the blue curve to high redshift data ($z>1$) and the black curve to the full dataset.}
\end{figure*}

Finally, we briefly consider the case where the parameters $R$ and $S$ are allowed to vary and then marginalized. Here there is one practical difficulty: given the lack on detailed knowledge of the physics of unification, it is difficult to identify a physically motivated choice of priors for both parameters. With this caveat, we agnostically choose broad uniform priors in the ranges $R=[-300,+300]$ and $S=[-1000,+1000]$. The results of this analysis are summarized in Figure \ref{fig6} and in Table \ref{table3}. It should be noted that the likelihood is highly non-Gaussian, so we provide both the range of parameters containing $68.3\%$ of the likelihood and the range for which the posterior $\Delta\chi^2$ is within $\Delta\chi^2\le4$ of the minimum.

\begin{table}
\begin{center}
\caption{Constraints on the relative variation of $\alpha$ for the low and high redshift data as well as for the full dataset, and on its current drift, when the phenomenological parameters $R$ and $S$ are marginalized. We show both the ranges which are within  $\Delta\chi^2\le4$ of the best-fit values and the range containing $68.3\%$ of the posterior likelihood (with the priors described in the text). Note that this posterior likelihood is highly non-Gaussian, as can be seen in Figure \ref{fig6}. All constraints are in ppm units.}
\label{table3}
\begin{tabular}{| l | c | c |}
\hline
Parameter & Range $\Delta\chi^2\le4$ & Range $68.3\%$ \\
\hline
$\frac{d\ln{\alpha}}{d(H_0t)}$ & $[-0.77, 0.39]$ & $0.01^{+0.43}_{-0.48}$ \\
\hline
$\frac{\Delta\alpha}{\alpha}$ ($z<1$) & $[-0.07, 0.05]$ & $-0.01^{+0.19}_{-0.22}$  \\
$\frac{\Delta\alpha}{\alpha}$ ($z>1$) & $[-0.04, 0.03]$ & $0.00^{+0.13}_{-0.96}$  \\
$\frac{\Delta\alpha}{\alpha}$ (All redshifts) & $[-1.17, 0.04]$ & $-0.01^{+0.12}_{-0.36}$ \\
\hline
\end{tabular}
\end{center}
\end{table}

In this case the null result (corresponding to no variations) is the preferred one. The reason for this is that within this parameter space one can find particular combinations of $R$ and $S$ that would lead to very large variations of $\mu$ and/or $g_p$, which would be excluded by the corresponding measurements unless $\alpha$ itself does not vary.

\section{Conclusions}

Tests of the stability of nature's fundamental constants are a powerful probe of the standard paradigms of cosmology and fundamental physics, and have the potential to identify the new physics which must be behind the theoretical pillars of modern cosmology---inflation, dark matter and especially dark energy. A plethora of these tests have been carried out in the last two decades (see \cite{Uzan,ROPP} for recent reviews) in an range of local, astrophysical and cosmological contexts. The ones which currently lead to the most sensitive tests are high-resolution spectroscopy of astrophysical systems (mainly in low-density absorption clouds along the line of sight of bright quasars) and laboratory comparisons of pairs of atomic clocks.

We have used standard chi-square analysis techniques to carry out a global analysis of all the currently available atomic clock and quasar data, studying both the consistency of tests of stability of different constants ($\alpha$, $\mu$ and $g_p$) and the consistency between local laboratory and astrophysical tests, under several alternative assumptions. This included both a model-independent analysis (studying the internal consistency of different types of measurements) but also model-dependent ones, for phenomenological motivated by string theory and grand unification.

Overall there is weak evidence of variations at up to three standard deviations, at the level of up to a few parts per million, and reasonable agreement between laboratory and astrophysical tests. The preferred values at low and high redshifts (in the acceleration and matter eras, respectively) are significantly different, though it's conceivable that these differences reflect (at least in part) the different systematics uncertainties affecting the low and high redshift measurement techniques which tend to be different. There are also concerns about systematics in the archival dataset of 293 $\alpha$ measurements of \cite{Webb} (see \cite{Syst} for a discussion), though our analysis indicates that the overall results are relatively unchanged whether one uses all available data or only the smaller dataset of 71 stringent dedicated measurements.

When it comes to specific models our analysis is limited by the lack of quantitative predictions for the relative variations of $\alpha$, $\mu$ and $g_p$ in string theory or unification scenarios. We have used the phenomenological parametrization of \cite{Coc,Luo} and considered particular examples of the two paradigms, as well as the particular case where only $\alpha$ varies while $\mu$ and $g_p$ do not. Interestingly, we find that among the three models the one in better agreement with all the data is the last one. In the future it will be important to obtain more quantitative predictions for specific classes of theoretically motivated models.

Meanwhile, the arrival of a new generation of high-resolution ultra-stable spectrographs such as ESPRESSO \cite{ESPRESSO} will significantly improve the sensitivity and reliability of these measurements, and should be able to confirm or rule out the current mild evidence for variations. Whether the outcome is one or the other, there will be significant implications for cosmology and fundamental physics.

\section*{Acknowledgements}

We are grateful to Ana Catarina Leite for helpful discussions on the subject of this work. This work was financed by FEDER---Fundo Europeu de Desenvolvimento Regional funds through the COMPETE 2020---Operacional Programme for Competitiveness and Internationalisation (POCI), and by Portuguese funds through FCT---Funda\c c\~ao para a Ci\^encia e a Tecnologia in the framework of the project POCI-01-0145-FEDER-028987. M.V.M. acknowledges financial support from Programa Joves i Ci\`encia, funded by Fundaci\'o Catalunya-La Pedrera (Spain).

\bibliographystyle{model1-num-names}
\bibliography{alpha}
\end{document}